\documentclass[journal,draftcls,onecolumn,12pt]{IEEEtranTCOM}
\usepackage[T1]{fontenc}
\usepackage[utf8]{inputenc}
\usepackage{amsmath} 
\usepackage{amsfonts} 
\usepackage{amssymb} 
\usepackage{amsthm}
\usepackage{color}
\usepackage{url}
\usepackage{caption}
\usepackage{subcaption}
\usepackage[linesnumbered,ruled,vlined]{algorithm2e}
\usepackage{setspace} 
\usepackage{verbatim}
\usepackage{multirow}
\usepackage{graphicx}
\usepackage{comment}

\SetKwInput{KwInput}{Input}                
\SetKwInput{KwOutput}{Output}              
\SetKwInput{Kwinitialize}{Initialization}              

\theoremstyle{remark}

\usepackage{cite}
\renewcommand{\vec}[1]{{\bf{#1}}} 
\newcommand{\vecgreek}[1]{{\boldsymbol{#1}}} 
\newcommand{\tran}{^{\mbox{\scriptsize T}}}
\newcommand{\herm}{^{\mbox{\scriptsize H}}}

\newcommand{\ex}[2][3]{^{\backslash #2 }}

\newcommand{\fro}[1]{\Vert #1\Vert_\mathrm{F}^2}
\newcommand{\norm}[1]{\Vert #1\Vert_2}

\newcommand{\gammab}{\vecgreek{\gamma}}

\newcommand{\Sigmab}{\vec{\Sigma}}
\newcommand{\Lambdab}{\vec{\Lambda}}
\newcommand{\Xib}{\vecgreek{\Xi}}

\newcommand{\Psib}{\vecgreek{\Psi}}

\newcommand{\mybibliography}{\bibliography{conf_short,jour_short,Bi}}

\def\red{\textcolor{red}}

\begin{document}
\title{Hierarchical MTC User Activity Detection and Channel Estimation with Unknown Spatial Covariance}
\author{\IEEEauthorblockN{Hamza Djelouat,~\IEEEmembership{Student Member,~IEEE,} Mikko J. Sillanpää,  Markus Leinonen$^{\dagger}$,~\IEEEmembership{Member,~IEEE,}  and Markku Juntti,~\IEEEmembership{Fellow,~IEEE}}
\thanks{H. Djelouat and M. Juntti are  are with Centre for Wireless Communications -- Radio Technologies, FI-90014, University of Oulu, Finland. Mikko J. Sillanpää is with Research Unit of Mathematical Sciences, University of Oulu, Finland.  e-mail: \{hamza.djelouat,markus.leinonen,markku.juntti\}@oulu.fi.\\
This work has been financially supported in part by the Academy of Finland (grant 319485) and Academy of Finland 6G Flagship (grant 346208).  H. Djelouat  would like to acknowledge the support of Tauno Tönning Foundation.  The work of M.\ Leinonen$^{\dagger}$, deceased, was also financially supported by the Academy of Finland (grant 340171).}}
\maketitle

\begin{spacing}{1.4}
\begin{abstract}

This paper addresses the joint user identification and channel estimation  (JUICE) problem in  machine-type communications under  the  practical spatially correlated channels model  with unknown covariance matrices. Furthermore, we consider an MTC network with hierarchical user activity patterns following an event-triggered traffic mode. Therein the users are distributed over clusters with a structured sporadic activity behaviour that exhibits both cluster-level and intra-cluster sparsity patterns. To solve the JUICE problem,  we first leverage the concept of strong priors   and  propose a hierarchical-sparsity-inducing spike-and-slab prior to model  the structured sparse activity pattern. Subsequently, we derive a Bayesian inference scheme by coupling the expectation propagation (EP) algorithm with the expectation maximization (EM) framework. Second, we reformulate the  JUICE as a \emph{maximum a posteriori} (MAP) estimation problem and  propose a  computationally-efficient solution based on the alternating direction method of multipliers (ADMM). More precisely, we relax the strong spike-and-slab prior with a cluster-sparsity-promoting prior based on the long-sum penalty. We then derive an ADMM algorithm that solves the MAP problem through a sequence of closed-form updates. Numerical results highlight the significant performance significant gains obtained by the proposed algorithms, as well as their robustness against various  assumptions on the users sparse activity behaviour.

\end{abstract}
\end{spacing} 
\vspace{-5mm}

\section{Introduction}

Machine-type communications (MTC)   constitute one of the fundamental pillars in the current 5G cellular systems \cite{shahab2020grant}. In MTC networks, the base station (BS)  aims to provide connectivity to a massive number of low-cost energy-constrained devices, known as  user equipments (UEs).
In the practical MTC, the UEs exhibit sporadic activity, generate predominantly uplink traffic and transmit mainly short-packet data \cite{bockelmann2016massive}. Consequently, employing the conventional channel access protocols in MTC networks would result in a large signalling overhead \cite{shahab2020grant}. 
To address this issue,  grant-free access protocols have been introduced,  enabling the UEs to access the network without the need for scheduling requests in advance. Therefore, utilizing grant-free access in MTC would result in a low signalling overhead and an extended lifespan for the UEs.



To fully exploit the features of grant-free  protocols, the BS has to accurately   perform  the task of joint UE identification and channel estimation (JUICE). Subsequently, owing to  the sporadic nature of UEs traffic,  the JUICE problem has been widely formulated as a sparse recovery problem. Furthermore, as the BS antennas sense the same sparse  activity behaviour, the JUICE problem extends to the multiple measurements vector (MMV) setup. Solving the sparse recovery problem in MMV  can be achieved via, for instance, greedy algorithms such as simultaneous orthogonal matching pursuit (SOMP)  \cite{tropp2006algorithms}, mixed norm optimization approaches  \cite{steffens2018compact}, sparse Bayesian learning (SBL) \cite{wipf2007empirical}, and message passing algorithms\cite{ziniel2012efficient}.
\vspace{-.4cm}
\subsection{Related Work}

Numerous ongoing efforts to devise algorithms to solve the JUICE  are presented in the literature. For instance, Chen {\it et al.}   \cite{chen2018sparse}  addressed the JUICE   under two scenarios, with and without prior knowledge of the large-scale fading. For both scenarios, they evaluated analytically a JUICE solution based on the AMP algorithm. Furthermore, Liu {\it et al.}  \cite{liu2018massive}  provided asymptotic performance analysis for a Bayesian  AMP algorithm in terms of activity detection and channel estimation under the assumption of known large-scale fading.   Ke {\it et al.} \cite{ke2020compressive} considered an enhanced mobile broadband network, and investigated the performance of the generalized AMP algorithm that exploits the structured sparsity  in the multiple-input multiple-output (MIMO) channels. The authors in \cite{chen2019covariance, Fengler2021non} proposed JUICE solutions from a non-Bayesian perspective, where they formulated the JUICE as a maximum likelihood problem and solved it via coordinate-block descend \cite{chen2019covariance} and non-negative least-squares \cite{Fengler2021non}. 
Nonetheless, these works focused on the JUICE only under \emph{ uncorrelated fading channels}, which may not accurately represent practical situations as the MIMO channels are typically spatially correlated \cite{sanguinetti2019towards}. 
 In fact, JUICE solutions that are designed for uncorrelated channels may exhibit sensitivity to the correlation structures encountered in practical scenarios \cite{cui2020jointly}.  Thus, incorporating the spatial correlation  in designing the JUICE solutions is of utmost importance, and it is still in its infancy in the literature.

 Recently, few works have looked into the JUICE problem  under spatially correlated channels. For instance, Cheng {\it et al.} exploited \red{in \cite{cheng2020orthogonal}} both the spatial and temporal correlation of the propagation channels and proposed the orthogonal AMP algorithm to solve the JUICE problem.
Rajoriya {\it et al.}  \cite{rajoriya2023joint}  proposed a Bayesian solution that couples AMP and SBL to provide an algorithm that enjoys the low complexity of AMP and the good performance of SBL. Bai  {\it et al.}  \cite{Bai2022Activity} proposed a distributed AMP algorithm in cell-free MTC networks that aims to reduce the complexity of AMP by distributing the computation load over several access points. Moreover, we formulated the JUICE as an $\ell_{2,1}$-norm minimization problem \cite{Djelouat2020Joint,Djelouat-Leinonen-Juntti-21-icassp} and as a \emph{ maximum a posteriori} (MAP) problem in \cite{djelouat2021spatial}. For both formulations, we derived computationally-efficient algorithms based on the alternating direction method of multipliers (ADMM). Furthermore, several theoretical analysis studies have investigated the performance of AMP for JUICE in correlated MIMO channels, in terms of activity detection accuracy \cite{Djelouat2021user, Jiang2022Performance}, channel estimation \cite {Jiang2022Performance} and  achievable rate \cite{rajoriya2023joint}.

While the works in \cite{cheng2020orthogonal,Djelouat2020Joint,Djelouat-Leinonen-Juntti-21-icassp,djelouat2021spatial,rajoriya2023joint,Bai2022Activity,Djelouat2021user,Jiang2022Performance} address the JUICE problem under the more practical spatially correlated MIMO channels, they make the  assumption that the  channel distribution information (CDI) for all the UEs are \textit{fully known to the BS at any transmission instance}. However, this assumption  can be challenging to fulfill in  realistic scenarios, as the BS cannot track the CDI of the UEs with long inactive status. Furthermore, the prior works on the JUICE  consider MTC networks with an independent   activity pattern amongst the UEs, modeling for instance,  a scenario where each UE monitor an independent processes and thus activate randomly.  Nonetheless, in practice, the  UEs are deployed over  several clusters, where the UEs within each cluster    
monitor the same phenomena, for example, leading to the paradigm of \emph{event-triggered traffic} models.


The event-triggered traffic model results in a \emph{hierarchical}  sparse activation  pattern constituting in both  \emph{cluster-level}  sparsity and \emph{intra-cluster} sparsity. The cluster-level  sparsity arises because the event epicenters are concentrated around a small subset of clusters (referred to as active clusters), causing only the UEs from those active clusters to be prompt for activity. On the other hand,  each event would trigger only a subset of UEs to be active in practice, resulting in an intra-cluster sparsity model\footnote{This can occur  if the UEs in  the same  cluster do not preform the same task, or when some of the UEs are not triggered by the event, in the case where they are far from the event epicentre.}. To the best of our knowledge, 
only two works have investigated the JUICE problem with correlated UEs activity pattern. Liu {\it et al.} \cite{Liu2022joint} addressed only the activity detection problem and proposed a  solution based on preamble selection under different assumptions on the prior knowledge on the activity correlation distribution. Becirovic {\it et al.}  \cite{Becirovic2023Activity}  proposed two sparsity-promoting priors based on $\ell_{2,1}$-norm and total-variation and proposed a non-negative least squares algorithm to solve a relaxed version of an $\ell_{0}-$norm minimization. 

\subsection{Main Contribution}


This paper makes the following two distinctions from the prior works: first, we address the JUICE in  \emph{spatially correlated} MIMO channels with \emph{no prior} knowledge on the exact CDI. Second,  in contrast to the  mainstream grant-free access literature, where the
traffic is assumed to be  independent amongst the UEs, 
we consider an MTC network where the UEs are distributed into clusters with an  activation pattern following an event-triggered traffic model, thus, introducing a \emph{correlated}  activity pattern amongst the UEs. For instance, this could model a network where the UEs form clusters based on their geographical locations and each cluster is associated with a common task. Here, an event could trigger a \emph{small subset} of UEs belonging to a particular cluster to activate concurrently, leading to hierarchical  UE activity pattern. 

The main \textbf{contributions} of our paper can be summarized as follows\footnote{Preliminary results of
this paper appear in \cite{Djelouat2023joint}.}:
\begin{itemize}
\item This paper proposes two Bayesian inference approaches to solve the JUICE problem in spatially correlated MIMO channels with a hierarchical  UEs activity pattern and a limited or no prior knowledge of the CDI. 
\item We propose a  \emph{hierarchical spike-and-slab} prior that incorporates  both the cluster-level and intra-cluster sparsity patterns. Subsequently, we derive a solution based on introducing an expectation propagation (EP) algorithm \cite{minka2001expectaion} within an expectation maximization (EM) \cite{bishop2006pattern} framework. The proposed solution iteratively approximates an intractable joint posterior distribution and provides an estimate on the active UEs, their channels, as well as the CDI.



\item We propose an alternative solution to the JUICE problem by relaxing the spike-and-slab prior with a log-sum prior \cite{candes2008enhancing} that  provides a more flexible approach to encode the hierarchical activity pattern. Subsequently, we formulate the JUICE as a MAP  problem and derive a solution based on the ADMM framework in order to solve iteratively an approximated version of the MAP problem via a  sequence of closed-form updates. 
\item Numerical results demonstrate the substantial performance gains achieved by our proposed algorithms, as well as their robustness against different activity pattern structures. 


 \end{itemize}

\section{JUICE with Correlated Activity and Partially Known CDI}

\subsection{System Model}
\begin{figure}
    \centering
    \includegraphics[scale=.4]{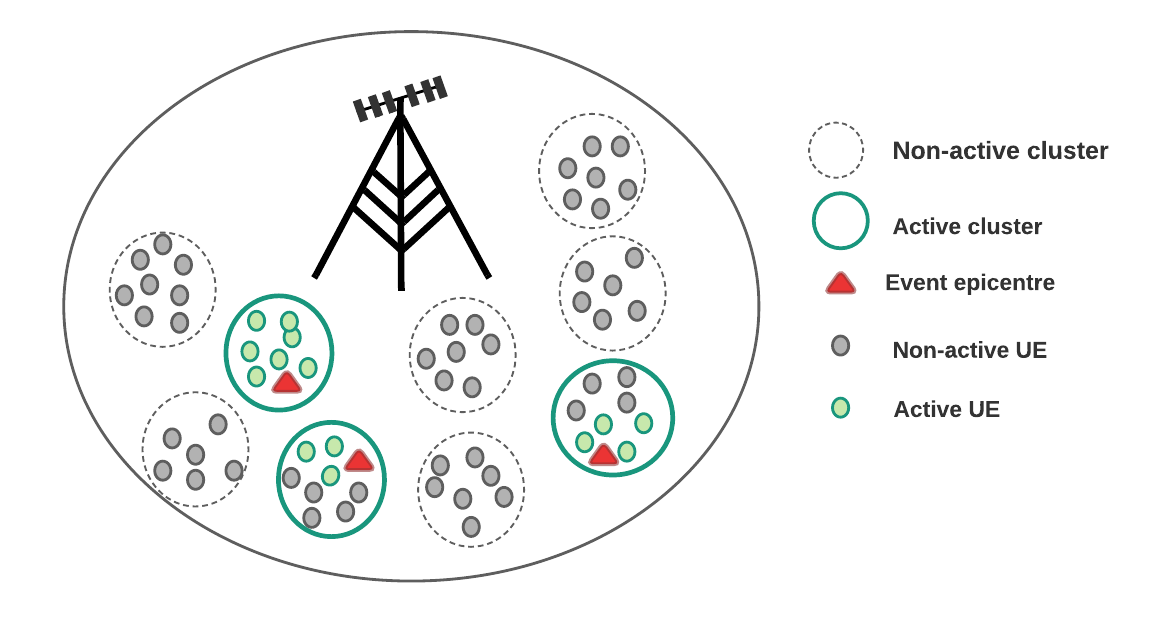}
    \caption{Illustration of an MTC network with an activity pattern centred around few number of clusters.}
    \label{fig::MTC_correlated}
\end{figure}

We consider a single-cell uplink network consisting of a set $\mathcal{N}=\{1,2,\ldots,N \}$  UEs served by a single BS equipped with  a  uniform linear array containing $M$ antennas, as depicted in Fig.\ \ref{fig::MTC_correlated}. The UEs are geographically distributed over $N_\mathrm{c}$ \textit{clusters}, where each UE  belong to a unique cluster. We denote the index set of the $l$th  cluster as  ${\mathcal{C}_{l}\subseteq\ \mathcal{N}}$,  ${l=1,\ldots,N_{\mathrm{c}}}$, where  each cluster contains $L$ UEs\footnote{For the simplicity of presentation, yet without loss of generality, we assume that all clusters contain the same number of UEs.}, such that ${N=LN_\mathrm{c}}$.

\vspace{-.1cm}

In contrast to the majority of the literature on grant-free access  that considers an independent UEs activation pattern,  we consider in this paper an MTC network operating in an event-triggered traffic model inducing a \emph{correlated activation pattern} between the UEs. In particular, event-triggered traffic model raises in the practical 
MTC network where the UEs are grouped into clusters, such that each  cluster is associated with a monitoring task. Subsequently, 
we make herein the following \emph{distinctions} based on  technical  observations on the UEs activation pattern under the event-triggered traffic model:
\begin{itemize}
    \item The UEs activity is triggered by events concentrated around a  small subset of \emph{active} clusters, thus, giving rise to a \emph{cluster-level} sparsity structure. Therefore, we define  the cluster-level  vector ${\vec{c}=[c_1,\ldots,N_{\mathrm{c}}]\tran}$, where $c_l=1$ if the $l$th cluster is active and $c_l=0$, otherwise.
    \item Within each active cluster, a subset  containing at most ${L_\mathrm{c}\leq L}$ UEs will be active,  thus, inducing a correlation between the UEs activity within the same cluster, in the form of \emph{intra-cluster} sparsity.  Thus, to model the intra-cluster sparsity, we introduce  $\gamma_i \in\{0,1\}$, $i \in \mathcal{N}$,  where $\gamma_i=1$ if the $i$th UE is active and $\gamma_i=0$, otherwise. Note that intra-cluster sparsity  imposes a special structure on $\gammab=[\underbrace{\gamma_1,\ldots,\gamma_L}_{\footnotesize\mbox{Cluster~1} },\underbrace{\gamma_{L+1},\ldots,\gamma_{2L}}_{\footnotesize\mbox{Cluster~2}},\ldots,\underbrace{\gammab_{N-L+1},\ldots,\gammab_{N}}_{\footnotesize\mbox{Cluster~}C}]$ where the elements belonging to a given cluster ($\gamma_i$, $i\in\mathcal{C}_l$) are assumed to be correlated. The correlation in the UES activity will be discussed in the following sections.
\end{itemize}

We consider that the  channel response $\vec{h}_i$ between the $i$th UE and the BS  follows a local scattering model \cite{massivemimobook}. Thus,  $\vec{h}_i \in \mathbb{C}^{M}$, $ \forall i \in \mathcal{N}$, is modeled  as a zero-mean complex Gaussian random variable, i.e., ${\vec{h}_i \sim  \mathcal{CN}(0,\vec{R}_i)}$,  such that $\vec{R}_i \in  \mathbb{C}^{M \times M}$ represents  the channel covariance matrix computed as ${\vec{R}_i=\mathbb{E}[\vec{h}_i\vec{h}_i\herm]}$. Additionally, we adopt the common assumption that the channels are  wide-sense stationary. Thus,  the changes in the    covariance matrices ${\vec{R}=\{\vec{R}_i\}_{i=1}^N}$ occur less frequently       compared to the variations in the channel realizations \cite{sanguinetti2019towards}. 

At any coherence interval  $T_{\mathrm{c}}$, each active UE transmits a total of    $\tau_{\mathrm{c}}$ symbols to the BS over two phases. In the first phase,   each active UE transmits a $\tau_{\mathrm{p}}$-length pilot sequence to  the BS, whereas in the  second phase,   each active UE transmits its information data to the BS using the remaining $\tau_{\mathrm{c}}-\tau_{\mathrm{p}}$ symbols. Subsequently, in order to decode the information data transmitted from the active UEs, the BS utilizes the received signal from the  pilot transmission phase to perform the JUICE task. To this end,  the pilot transmission phase is realized by assigning to each UE $i$, $\forall i \in \mathcal{N}$, a unique unit-norm pilot sequence $\vecgreek{\phi}_i \in \mathbb{C}^{\tau_{\mathrm{p}}}$, and  a  transmit power $p_i$ inversely proportional to its average channel gain in order   to 
to reduce the disparity in the channels gain amongst the UEs. \cite{Marata2023joint,bjornson2016massive}. 
Consequently, the received  signal during  the   pilot transmission phase $\vec{Y} \in \mathbb{C}^{\tau_{\mathrm{p}}\times M}$ is given by
\begin{equation}
\label{eq::Y}
\vec{Y}=\sum_{i=1}^{N}\gamma_i \sqrt{p_i}   \vecgreek{\phi}_i\vec{h}_i\tran+\vec{W}=\vec{\Phi} \vec{X}\tran+ \vec{W},
\end{equation}
where  ${\vec{X}=[\vec{x}_1,\ldots,\vec{x}_{N}] \in \mathbb{C}^{M\times N}}$ represents  the effective channel matrix with ${\vec{x}_i=\gamma_i \sqrt{p_i} \vec{h}_i}$, $\vec{\Phi}=[\vecgreek{\phi}_1,\ldots,\vecgreek{\phi}_N] \in \mathbb{C}^{\tau_{\mathrm{p}}\times N}$ is the  pilot sequence matrix , and $\vec{W}\in\mathbb{C}^{\tau_{\mathrm{p}}\times M} $ is an additive white Gaussian noise with an i.i.d.\ elements, where  each element is drawn from  $\mathcal{CN}(0,\,\sigma^{2})$.

The joint detection of the active UEs and estimating their channel reduces to estimating the (unknown) \emph{row-sparse} effective channel matrix $\vec{X}\tran$ based on the received pilot signal  in \eqref{eq::Y}. Thus, the JUICE can be formulated as a sparse recovery problem from an MMV setup. The prior works in the literature showed that sparse recovery algorithms derived from a Bayesian perspective yield usually the best performance  \cite{ziniel2012efficient}. 

\subsection{ Bayesian Inference Setup}
 In the Bayesian framework,   the unknown variables to be estimated, i.e., $\vec{X}$, is modelled   using a prior probability distribution functions (PDF) which incorporate and encode  both the prior knowledge on $\vec{X}$ as well as the prior knowledge on its hidden hyper-parameters. Subsequently,  utilizing prior functions that promote sparsity while incorporating the  hierarchical sparse structures of $\vec{X}\tran$ is the key to achieve accurate solutions for the JUICE problem. In particular,  sparsity-promoting priors can be  categorized as either weak or strong priors. For instance,   the Laplace distributions is a  weak sparsity prior as it promotes sparsity, but it does not assign strictly a zero probability to the case where the vector $\vec{x}_i=0$. On the other hand,  strong sparsity priors, such as the spike-and-slab distributions (with delta peak at zero), are more stringent as they assign a strictly positive probability to the case of $\vec{x}_i=0$ \cite{andersen2014bayesian}. 
Therefore, for the  JUICE problem in this paper,  the priors should be designed in order to the encode (i) the spatial correlation of each $\vec{x}_i$, $\forall i$, (ii) the cluster-level  sparsity between clusters, (iii) the intra-cluster  sparsity structure. We will discuss next, how to design the priors  to encode (i), (ii), and (iii). 

\subsection{Hierarchical Spike-and-Slab Prior}     
In the following, we leverage the concept of strong sparsity priors and we propose a \emph{hierarchical spike-and-slab sparsity-promoting prior} to model the cluster-level sparsity, the intra-cluster sparsity, and the spatially correlated channels. 
First, we introduce the following parameters:
\paragraph{Cluster-level activity prior }
In order to impose the cluster-level activity we model each $c_l$ as a  Bernoulli random variable with ${p(c_l=1) =\epsilon}$ and ${p(c_l=0) = 1-\epsilon}$. Furthermore, to account for the   independence amongst the clusters activity, we express the probability mass function of $ p(\vec{c})=   \prod_{l=1}^{N_\mathrm{c}}\mathcal{B}(c_l;\epsilon) $, where $\mathcal{B}$ denotes the Bernoulli distribution.
\
\paragraph{Intra-cluster sparsity prior}
 We define the hyper-parameters  ${\gamma_i} \in \mathbf{R}^{+}$, $\forall i$, that imposes a row sparsity structure on $\vec{X}$. To this end, we design $p(\bar{\gammab})$ such that it promotes sparse solution, for instance,  $p(\bar{\gammab})$ can be drawn from the Laplacian distribution as $p(\bar{\gammab})\propto \exp\big( -\textstyle\sum_{i=1}^{N}\bar{\gamma_i} \big)$.

\paragraph{ Channels spatial correlation prior}
we introduce the set of covariance matrices $\{\bar{\vec{R}}_i\}_{i=1}^N$, such that each $\bar{\vec{R}}_i$ is positive definite matrix  that captures the spatial correlation between the entries of $i$th row in $\vec{X}\tran$.   
A common and physically grounded prior for an   $\bar{\vec{R}}_i$   of the Gaussian random variable $\vec{x}_i$  is given by the inverse Wishart distribution \cite{bishop2006pattern}, defined as
\vspace{-.4cm}\begin{equation}\label{eq:I-Wishart}
        p(\bar{\vec{R}}_i)\sim \mathcal{IW}(\bar{\vec{R}}_i;\vec{B}_i,v)= f(\vec{B}_i,v)  |\Sigmab_i|^{-d} \exp\left[-\text{Tr}(\vec{B}_i \bar{\vec{R}}_i^{-1})\right]
\end{equation}
where $f(\vec{B}_i,v)$ is a normalization constant, ${d=v-M+1>0}$, $v$  controls the degrees of freedom of the distribution, and ${\vec{B}_i \in \mathbb{C}^{M \times M}}$ is a symmetric positive-definite  matrix that represents the prior information for the covariance matrix $\bar{\vec{R}}_i$ \cite{bishop2006pattern}.

By utilising the definitions above, we  model the effective channel $\vec{x}_i$, $ \forall i \in \mathcal{N}$, using the spike-and-slab prior as
\vspace{-.4cm}\begin{equation}\label{spike-slab}
     p(\vec{x}_i|c_l,\bar{\gamma_i}, \bar{\vec{R}}_i)=(1-c_l)\delta(\vec{x}_i)+c_l \mathcal{CN}(\vec{x}_i;\vec{0},\bar{\gamma}_i\bar{\vec{R}}_i).
\end{equation}
The main idea in \eqref{spike-slab} can be summarized as follows:
\begin{itemize}
    \item If $c_l=0$, the vector $\vec{x}_i$ would have only the spike component, i.\ e.\ , delta function, from \eqref{spike-slab}, thus leading to an estimation of a zero-vector  ($\vec{x}_i=\vec{0}$).
    \item  If $c_l=1$, $\vec{x}_i$ would have only the slab component from \eqref{spike-slab} in the form be a Gaussian random vector with covariance matrix $\bar{\gamma_i}\bar{\vec{R}}_i$. Therefore,  if  $\bar{\gamma}_i\approx 0$,  the covariance matrix $\bar{\gamma}_i\bar{\vec{R}}_i$  of the  slab component in \eqref{spike-slab}  would be very small  that we could safely estimate that $\vec{x}_i \approx \vec{0}$, whereas if $\bar{\gamma}_i>0$, $\vec{x}_i$ would be a non-zero Gaussian random vector.
\end{itemize}

Finally,  to encode (ii)-(iii) into the Bayesian formulation, we propose the following \emph{Hierarchical spike and slab prior} on  $p(\vec{X}|\vec{c},\vec{R})$ as 
\begin{equation}\label{eq:P(X)_l}
    \displaystyle p(\vec{X}|\vec{c},\vec{R})=\prod_{l=1}^{N_\mathrm{c}} p(\vec{X}_{\mathcal{C}_l}|c_l)  =\prod_{l=1}^{N_\mathrm{c}} \bigg[(1-c_l)\delta\big(\vec{X}_{\mathcal{C}_l}\big)+c_l\prod_{i \in {\mathcal{C}_l}} \mathcal{CN}(\vec{x}_i;\vec{0},\vec{R}_i) \bigg], \\
  \end{equation}
where $\delta\big(\vec{X}_{\mathcal{C}_l}\big)=\prod_{i \in {\mathcal{C}_l}}\delta(\vec{x}_i)$, and we denoted for simplicity $\vec{R}_i=\bar{\gamma}_i\bar{\vec{R}}_i$.

\section{JUICE Via EM-EP}\label{sec::EM-EP}
A common approach in Bayesian inference when the hyper-parameter set $\vecgreek{\Xi}=\{\bar{\gamma_i},\bar{\vec{R}}_i\}_{i=1}^N$  is not known, is  to maximize the likelihood function $p(\vec{Y}|\vecgreek{\Xi})$. However, in many cases, the likelihood $p(\vec{Y}|\Xib)$  is a non-convex function of $\Xib$ and its global maximum cannot be found in a closed form. Thus, $\Xib$ can be obtained  through type-II maximum likelihood estimation by finding a local maximum using  the expectation-maximization (EM) framework.  The classical EM  iteratively alternates between two steps, namely, the E-step and the M-step. In the E-step, the current values of the hyper-parameters are used to evaluate the posterior distribution of interest. Subsequently, the hyper-parameters are estimated again using the   current statistics of the posterior distribution in the M-step \cite{bishop2006pattern}.

In the JUICE context, for a given $\Xib$, the joint posterior distribution $p(\vec{X},\vec{c}|\vec{Y})$ is expressed as a product for three factors, namely, $f_1(\vec{X})$, $f_2(\vec{X},\vec{c})$ and $f_3(\vec{c})$ as
  \begin{equation}
\begin{array}{ll}\label{eq::posterior}f(\vec{X},\vec{c})&\displaystyle=p(\vec{X},\vec{c}|\vec{Y})\overset{}{=}\frac{1}{p(\vec{Y})}{p(\vec{Y}|\vec{X})p(\vec{X}|\vec{c})p(\vec{c})}\\
&=~\dfrac{1}{p(\vec{Y})}\underbrace{\mathcal{CN}(\vecgreek{\Phi}\vec{X}\tran,\sigma^2\vec{I}_m)}_{f_1(\vec{X})}\underbrace{\prod_{l=1}^{N_{\mathrm{c}}}\big[(1-c_l)\delta\big(\vec{X}_{\mathcal{C}_l}\big)+c_l\prod_{i \in {\mathcal{C}_l}} \mathcal{CN}(\vec{x}_i;\vec{0},\vec{R}_i) \big]}_{f_2(\vec{X},\vec{c})}\underbrace{\prod_{l=1}^{N_{\mathrm{c}}}\mathcal{B}(\vec{c}_l;\epsilon)}_{f_3(\vec{c})}.
\end{array}
\end{equation} 
The  disadvantage of the spike-and-slab prior is that it renders the computation of the posterior distribution  to be a computationally demanding Task. In particular,  $p(\vec{X},\vec{c}|\vec{Y})$ in \eqref{eq::posterior} cannot be computed exactly when $N$ is large and, thus, it  has to be estimated numerically. To this end, we  resort to the expectation propagation (EP) algorithm to find a tractable approximation to the true posterior  distribution $p(\vec{X},\vec{c}|\vec{Y})$ in  
 \eqref{eq::posterior}.

Next, we shall derive a novel JUICE solution  based on coupling the EP algortihm within the  EM framework.  More precisely, at any EM iteration $(k)$:  (i)  given the set of hyper-parameter $\Xib^{(k-1)}$, the EP framework is utilized to approximate the intractable posterior distribution $p(\vec{X},\vec{c}|\vec{Y})$, and subsequently, compute the posterior mean, denoted as $\vec{m}_i^{(k-1)}$, and covariance matrix, denoted as $\Sigmab_i^{(k-1)}$, of the effective channel $\vec{x}_i$,  $\forall i \in \mathcal{N}$. (ii) In the M-step, by utilizing the sufficient statistics of $\vec{X}$ obtained from the previous E-step, we compute the new values of the hyper-parameters $\Xib^{(k)}$ by minimizing a lower-bound on the negative
 log-likelihood $p(\vec{Y}|\vecgreek{\Xi})$. Next, we will present  the details of the proposed algorithm. 
\vspace{-.4cm}\subsection{Main Idea of EP}
 The EP algorithm aims to approximate iteratively the true posterior distribution $f(\vec{X},\vec{c})$ by a simpler distribution $Q(\vec{X},\vec{c})$ that belongs to an exponential family. More precisely,  EP aims is to approximate the factors $f_1(\cdot)$, $f_2(\cdot)$, $f_3(\cdot)$ in \eqref{eq::posterior} by $q_1(\cdot)$, $q_2(\cdot)$, $q_3(\cdot)$, respectively, such that 
\vspace{-.5cm}\begin{equation}
   f(\vec{X},\vec{c})\approx Q(\vec{X},\vec{c})=\frac{1}{K^{\mathrm{EP}}}q_1(\vec{X})q_2(\vec{X},\vec{c})q_3(\vec{c}).
\end{equation}

Each  factor $q_k(\cdot)$, $k=1,2,3$, of the joint variations approximation $Q(\vec{X},\vec{c})$ is obtained by minimizing iteratively the Kullback-Leibler divergence \cite{bishop2006pattern}  as
\begin{equation}\label{eq:minKL}
    q_k^{*}=\min_{q_k} \mathrm{KL}\bigg( f_k(\cdot)Q\ex{k}(\cdot)\;\Big|\Big| \;q_k(\cdot)Q\ex{k}(\cdot)\bigg), k=1,2,3,
\end{equation}
where $Q\ex{k}(\cdot)=\frac{Q(\cdot)}{q_k(\cdot)}$ is termed the cavity distribution. The optimization problem \eqref{eq:minKL} is convex with a unique global optimum solution, that is obtained  by matching the expected values of the sufficient statistics of $q_k(\cdot)Q\ex{k}(\cdot)$ to the ones of $f_k(\cdot)Q\ex{k}(\cdot)$ \cite{bishop2006pattern}. In the following section, we show in details how to  derive the approximation factor $Q(\vec{X},\vec{c})$ through the EP framework.

\vspace{-.5cm}\subsection{E-Step: Posterior approximation Via EP}
The choice of the approximating factors in EP is not stringent, rather, it  is flexible. Thus, we design $q_1(\cdot)$, $q_2(\cdot)$, $q_3(\cdot)$ such that they: 1) offer  tractability and closed-from updates, 2) capture the important features of the true posterior distribution such as cluster-level and intra-cluster sparsity. To this end, we design the approximate factors as follows
\vspace{-.3cm}\begin{equation}
q_1(\vec{X})=\prod_{i=1}^Nq_1(\vec{x}_i)=\prod_{i=1}^N\mathcal{CN}(\vec{x}_i;\vec{m}_{1,i},\Sigmab_{1,i}),
\end{equation}\vspace{-.6cm}
\begin{equation}\label{q_2}
q_2(\vec{X},\vec{c})=\prod_{l=1}^{N_\mathrm{c}} q_2(\vec{X}_{\mathcal{C}_l},c_l)=\prod_{l=1}^{N_\mathrm{c}}c_l\prod_{i \in \mathcal{C}_l}\mathcal{CN}(\vec{x}_i;\vec{m}_{2,i},\Sigmab_{2,i}), 
\end{equation}\vspace{-.8cm}
\begin{equation}
q(\vec{c})=\prod_{l=1}^{N_\mathrm{c}}\mathcal{B}(c_l;\epsilon).
\end{equation}
Thus, we write the global approximation $Q(\vec{X},\gammab,\vec{c})$ as 
\begin{equation}\label{eq:Q}\vspace{-.6cm}
\begin{array}{ll}
Q(\vec{X},\gammab,\vec{c})&\displaystyle\propto
q_1(\vec{X})q_2(\vec{X},\vec{c})q_3(\vec{c})\propto\prod_{i=1}^N \mathcal{CN}(\vec{x}_i;\vec{m}_{1,i},\Sigmab_{1,i})  \mathcal{CN}(\vec{x}_i;\vec{m}_{2,i},\Sigmab_{2,i}) \prod_{l=1}^{N_\mathrm{c}}c_l\mathcal{B}(c_l;\epsilon) \\
     & \displaystyle\propto\prod_{i=1}^N\mathcal{CN}(\vec{x}_i;\vec{m}_i,\Sigmab_i) \prod_{l=1}^{N_\mathrm{c}}c_l\mathcal{B}(c_l;\epsilon),
\end{array}
\end{equation}
with
\begin{equation}\label{eq:Q_moments}
\begin{array}{ll}
  \Sigmab_i &=\big(\Sigmab_{1,i}^{-1}+\Sigmab_{2,i}^{-1}\big)^{-1} ,\quad \quad i=1,\cdots,N,\\
\vec{m}_i&=\Sigmab_i\big(\Sigmab_{1,i}^{-1}\vec{m}_{1,i}+\Sigmab_{2,i}^{-1}\vec{m}_{2,i}\big) ,\quad \quad i=1,\cdots,N,
\end{array}
\end{equation}
where  $\vec{m}=[\vec{m}_1,\ldots,\vec{m}_N]$ and  $\Sigmab=\{\Sigmab_i\}_{i=1}^N$ are obtained by applying the product of two Gaussian terms rule, as shown in Appendix \ref{app:2gauss}. 
We note that since $q_3(\vec{c})$ is the same as $f_3(\vec{x})$,  it can be obtained directly and we need to estimate only $q_1(\vec{X})$ and $q_2(\vec{X},\vec{c})$ as we show next.
\subsubsection{Estimation of $q_1(\vec{X})$}

We describe now how to  compute the  $\{\vec{m}_{1,i}, \Sigmab_{1,i}\}_{i=1}^N$ of the first approximate term   $q_1(\vec{X})$. Note that since both the $f_1(\vec{X})$ and $q_1(\vec{X})$ has a Gaussian form,  $f_1(\vec{X})$ can be approximated exactly by $q_1(\vec{X})$, independently of the values of the other approximate factors $q_2(\cdot)$ and $q_3(\cdot)$.  Subsequently,  we only have to set $q_1(\cdot)=f_1(\cdot)$ at the start of the EP algorithm and it can be kept constant afterwards.

Now, let us  rewrite the vector form of the received signal as
\vspace{-.5cm}\begin{equation}
    \vec{y}=\vec{\Theta}\bar{\vec{x}}+  \vec{w},
    \label{Y_vec}
 \end{equation}
where  $\vec{x}=\mathrm{vec}(\vec{X})$, ${ \vec{w}=\mathrm{vec}(\vec{W}\tran) \in \mathbb{C}^{ \tau_{\mathrm{p}}M}} $, ${\vec{y}=\mathrm{vec}(\vec{Y}\tran)\in \mathbb{C}^{\tau_{\mathrm{p}}M}}$,  and ${\vec{\Theta}=\vecgreek{\Phi}\otimes\vec{I}_{M} \in \mathbb{C}^{M\tau_{\mathrm{p}} \times NM}}$, where $\otimes$  denotes the Kronecker product and the operation $\mathrm{vec}(\cdot) $  stacks the columns of the matrix vertically. Subsequently the vector-form of the likelihood function $f_1(\cdot)$ is given by
\vspace{-.5cm}\begin{equation}\label{eq:f_1bar}
   f_1(\bar{\vec{x}})= p(\vec{y}|\bar{\vec{x}})\sim\mathcal{CN}(\vec{y};\Theta\bar{\vec{x}},\sigma^2\vec{I}_{M\tau_{\mathrm{p}}}).
\end{equation}
Similarly, we can write the vector form of $q_1(\vec{X})$ as 
\vspace{-.5cm}\begin{equation}\label{eq:q_1bar}
    q_1(\bar{\vec{x}})~\sim \mathcal{CN}(\bar{\vec{x}};\bar{\vec{m}}_1,\bar{\Sigmab}_1),
\end{equation}   
where   $ {\bar{\Sigmab}_1=\mathrm{diag}( \Sigmab_{1,1},\ldots,\Sigmab_{1,N})}$ is a block diagonal matrix and $\bar{\vec{m}}_1=[\vec{m}_{1,1}\tran,\ldots,\vec{m}_{1,N}\tran]\tran$. 
Note that $ f_1(\bar{\vec{x}})$ is a distribution of $\vec{y}$ conditioned on $\bar{\vec{x}}$, whereas  $q_1(\bar{\vec{x}})$ is a function of $\bar{\vec{x}}$ that depends on $\vec{y}$, $\bar{\vec{m}}$ and $\bar{\Sigmab}$. Thus, by writing the full Gaussian distributions in \eqref{eq:f_1bar} and \eqref{eq:q_1bar}, and rearranging few terms, the first and second moments of $q_1(\vec{\bar{x}})$ are parameterized as follows
\vspace{-.5cm}\begin{equation}
    \begin{array}{cc}
      \bar{\Sigmab}_1^{-1}  &= \frac{1}{\sigma^2}\vecgreek{\Theta}\herm\vecgreek{\Theta},\quad \quad\quad\quad    \bar{\Sigmab}_1^{-1}\vec{\bar{m}}_1=\frac{1}{\sigma^2}\vecgreek{\Theta}\herm\vec{y}.
    \end{array}
\end{equation}


\subsubsection{Estimation of  $q_2(\cdot)$}
Note that  $q_2(\vec{X},\vec{c})$ in \eqref{q_2} factorize into $N_c$  independent mixed Gaussian-Bernoulli distributions $q_2(\vec{X}_{\mathcal{C}_l},c_l)$, $l=1,\ldots,N_{\mathrm{c}}$, allowing for the parallel updates  of $q_2(\vec{X},\vec{c})$. In the following, we present in detail how to update each term $q_2(\vec{X}_{\mathcal{C}_l},c_l)$.

\paragraph{Update  $Q^{\backslash 2,l}(\cdot)$}
First, we compute the marginal cavity distribution $Q^{\backslash 2,l}(\vec{X}_{\mathcal{C}_l},c_l)$ by removing the contribution of $q_2(\vec{X}_{\mathcal{C}_l},\vec{c}_l)$ from the the global approximation $Q(\vec{X},\vec{c})$ in \eqref{eq:Q} as 
\vspace{-.3cm}\begin{equation}\label{Q_cavity}
    Q^{\backslash 2,l}(\vec{X}_{\mathcal{C}_l},c_l)=\frac{Q(\vec{X},\vec{c})}{q_2(\vec{X}_{\mathcal{C}_l},c_l)}\propto\prod_{i \in \mathcal{C}_l}\mathcal{CN}(\vec{x}_i;\hat{\vec{m}}_i\ex{2,l},\hat{\Sigmab}_i\ex{2,l})\mathcal{B}(c_l;\epsilon) ,
\end{equation}
where $\hat{\Sigmab}_i\ex{2,l}$ and $ \hat{\vec{m}}_i\ex{2,l}$ are obtained by utilizing the rule of the fraction of two  Gaussian terms, as shown in Appendix \ref{app:2gauss}, and they are given as   
\vspace{-.5cm}\begin{equation}\label{moments_hat}\begin{array}{ll}
     \hat{\Sigmab}_i\ex{2,l}&   =\big( \Sigmab_{i}^{-1}-\Sigmab_{2,i}^{-1}\big)^{-1}, \;\;\forall i \in \mathcal{C}_l,  \\
      \hat{\vec{m}}_i\ex{2,l}  & =  \hat{\Sigmab}_i\ex{2,l}\big( \Sigmab_{i}^{-1}\vec{m}_i-\Sigmab_{2,i}^{-1}\vec{m}_{2,i}\big), \;\;\forall i \in \mathcal{C}_l.
\end{array}
\end{equation}

\paragraph{Update $Q^{\mathrm{new}}$ } Next we update $Q^{\mathrm{new}(\cdot)}=q_2(\cdot)Q\ex{2,l}(\cdot)$ by minimizing a KL-divergence distance given as $\mathrm{KL}\bigg( \frac{1}{G_{l,0}}f_2(\cdot)Q\ex{2,l}(\cdot)\;\big|\big|\; q_2(\cdot)Q\ex{2,l}(\cdot)\bigg)$, where $G_{l,0}$ is the normalizing constant
needed to ensure that the $f_2(\cdot)Q\ex{2,l}(\cdot)$ integrates to unity, given as (Please refer to Appendix\ \ref{app::q_2})
\vspace{-.9cm}\begin{equation}\label{eq:G_0}
G_{l,0}= \displaystyle\sum_{\vec{c}}\int f_2(\vec{X}_{\mathcal{C}_l},c_l)Q\ex{2,l}(\vec{X}_{\mathcal{C}_l},\vec{c})d\vec{X}d\vec{c}=\displaystyle b_l +a_l,
\end{equation}
where $b_l=(1-\epsilon)\displaystyle\prod_{i\in \mathcal{C}_l}\mathcal{CN}(\vec{0};\hat{\vec{m}}_i\ex{2,l},\hat{\Sigmab}_i\ex{2,l})$ and $a_l=\epsilon\displaystyle\prod_{i\in \mathcal{C}_l} \mathcal{CN}(\vec{0};\hat{\vec{m}}_i\ex{2,l},\hat{\Sigmab}_i\ex{2,l}+\vec{R}_i)$.

Since all the terms in $Q(\cdot)$ and $f_2(\cdot)$ are drawn from a Gaussian distribution, the solution to the previous KL-divergence is given by matching the moments  of $Q^{\mathrm{new}}(\cdot)$ and $f_{2,l}(\cdot)Q\ex{2,l}(\cdot)$. To this end, we compute the sufficient statistics of $f_{2,l}(\cdot)Q\ex{2,l}(\cdot)$  with respect to  both  $c_l$ and $\vec{x}_i$, $i  \in \mathcal{C}_l$. Subsequently,  the sufficient central moments of $Q^{\mathrm{new}}(\cdot)$ are given as
\begin{equation}\label{q_new_moments}
    \begin{array}{ll}
    \mathbb{E}_{f_{2,l}Q\ex{2,l}}[c_l]&=\displaystyle\frac{a_l}{a_l+b_l}\\ 
          
          \mathbb{E}_{f_{2,l}Q\ex{2,l}}[\vec{x}_i]&=\displaystyle\frac{1}{ G_{l,0}}\big( a_l\vec{R}_i(\vec{R}_i+\hat{\Sigmab}_i\ex{2,l})^{-1}\hat{\vec{m}}_i\ex{2,l} \big)  \\
          \textrm{Var}_{f_{2,l}Q\ex{2,l}}[\vec{x}_i]&=\displaystyle\frac{a_l}{ G_{l,0}}\big(\hat{\Sigmab}_i\ex{2,l^{-1}}+\vec{R}_i^{-1}\big)^{-1}+\big(\frac{1}{G_{l,0}a_l}-1\big) \mathbb{E}_{f_{2,l}Q\ex{2,l}}[\vec{x}_i]  \mathbb{E}_{f_{2,l}Q\ex{2,l}}[\vec{x}_i]\herm,  \\
    \end{array}
\end{equation}
The details of the derivations to obtain \eqref{q_new_moments} are presented in Appendix \ref{app::q_2}.

\paragraph{Update $q_2(\vec{X}_{\mathcal{C}_l},c_l)$} finally, the updated $q_2(\cdot)$ is computed as
\begin{equation}\label{eq:q_new}
    q_2(\vec{X}_{\mathcal{C}_l},c_l)=\frac{Q^*(\vec{X},\vec{c})}{Q\ex{2,l}(\vec{X}_{\mathcal{C}_l},c_l)}\propto \prod_{i \in \mathcal{C}_l}\mathcal{CN}(\vec{x}_i;\vec{m}_{2,i},\Sigmab_{2,i}) \mathcal{B}\big(c_l|\frac{a_l}{a_l+b_l}\big)
\end{equation}
where the  mean and the covariance matrices of the updated $q_2(\vec{X}_{\mathcal{C}_l},c_l)$    are computed as 
\vspace{-.4cm}\begin{equation}\label{eq:Sigma2}
    \Sigmab_{2,i}=\big( \textrm{Var}_{f_{2,l}Q\ex{2,l}}[\vec{x}_i]^{-1}  -\hat{\Sigmab}_{2,i}\ex{{2,l}^{-1}}\big)^{-1}, i \in \mathcal{C}_l,
\end{equation}
\begin{equation}\label{eq:m_2}
    \vec{m}_{2,i}= \Sigmab_{2,i}^{-1}\big( \textrm{Var}_{f_{2,l}Q\ex{2,l}}[\vec{x}_i]^{-1}\mathbb{E}_{f_{2,l}Q\ex{2,l}}[\vec{x}_i] -\hat{\Sigmab}_{2,i}\ex{{2,l}^{-1}}\hat{\vec{m}}_{i}\ex{2,l}\big)^{-1}, i \in \mathcal{C}_l.
\end{equation}
\subsection{M-Step: Hyper-parameter Update }
Once   $q_2(\vec{X},\vec{c})$ is updated using the previous $\Xib^{(k-1)}$, the new posterior mean $\vec{m}^{(k)}$ and posterior covariance $\Sigmab^{(k)}$ of $Q(\vec{X},\vec{c})$ are updated using \eqref{eq:Q_moments}. Subsequently, the M-step at the $k$th EM iteration is carried out as follows
\begin{equation}\label{eq:Q_theta}
    \begin{array}{ll}
        \Xib^{(k)}&=\underset{\Xib}{\max}~\mathbb{E}_{\vec{X},\vec{c}|\vec{Y},\Xib^{(k-1)}} \log p(\vec{y},\vec{X}|\Xib) 
\overset{(a)}{=}\underset{\Xib}{\max}~\mathbb{E}_{\vec{X},\vec{c}|\vec{Y},\Xib^{(k-1)}}\big[\log p(\vec{X}|\vec{c},\bar{\gammab},\bar{\vec{R}})
+\log p(\bar{\gammab})+     \log p(\bar{\vec{R}})\big]\\ 

&\propto   \displaystyle\underset{\Xib}{\max}~
\sum_{i=1}^N -M \log(\bar{\gamma}_i)- |\bar{\vec{R}}_i|-\text{Tr} \bigg[\bar{\gamma}_i^{-1}\vec{R}_l^{-1}\mathbb{E}_{\vec{X},\vec{c}|\vec{Y},\Xib}[\vec{x}_i\vec{x}_i\herm]\bigg]
 - \log p(\bar{\gammab})+     \log p(\bar{\vec{R}})\\
&\propto   \displaystyle\underset{\Xib}{\max}~
\sum_{i=1}^N -M \log(\bar{\gamma}_i)- |\bar{\vec{R}}_i|- \text{Tr} \bigg[\gamma_i^{-1}\vec{R}_i^{-1}\big(\vec{m}_i^{(k)}\vec{m}_i^{(k)\herm}+\Sigmab_i^{(k)}\big)\bigg]
 - \log p(\bar{\gammab})+     \log p(\bar{\vec{R}}),\\

    \end{array}
\end{equation}
where $(a)$ is obtained by dropping the terms that do not depend on $\Xib$ from joint probability and noting that $p(\Xib)=p(\vec{\gammab})p(\bar{\vec{R}})$  by design. 


\subsubsection{$\bar{\gammab}$-update}
Note that the final expression in \eqref{eq:Q_theta} shows that the   prior $p(\bar{\gammab})$ plays a role in the M-step. However, it has  been shown that even a non-uniform prior will lead to a sparse vector $\bar{\gammab}$, thanks to the relevance vector machine \cite{tipping2001sparse}. Thus we will make the same simplification and drop $p(\bar{\gammab})$ from \eqref{eq:Q_theta}. Subsequently,  \eqref{eq:Q_theta}  decouples  into $N$ subproblem across $\bar{\gammab}$ as 
\vspace{-.5cm}\begin{equation}\label{eq:gamma_bar_opt}
    \bar{\gamma_i}^{(k)}= \displaystyle\underset{\bar{\gamma}_i \geq0}{\max}~
-M \log(\bar{\gamma}_i)-  \text{Tr} \bigg[\bar{\gamma}_i^{-1}\vec{R}_i^{-1}\big(\vec{m}_i^{(k)}\vec{m}_i^{(k)\herm}+\Sigmab_i^{(k)}\big)\bigg], \quad \forall i.
\end{equation}
Thus, the optimal solution is obtained by setting the gradients of the objective function in \eqref{eq:gamma_bar_opt} with respect to $\bar{\gamma}_i$ to zero, resulting in the following update rule
\begin{equation}\label{eq:gamma_bar_sol}
\bar{\gamma}_i^{(k)}=\frac{1}{M}\text{Tr}\bigg[\bar{\vec{R}}_i^{-1}\big(\vec{m}_i^{(k)}\vec{m}_i^{(k)\herm}+\Sigmab_i^{(k)}\big)\bigg], \quad \forall i.
\end{equation}
\subsubsection{Covariance matrix $\{\bar{\vec{R}}_i\}$-update}\label{R_update_EP}

Taking inspiration from the argument presented in \cite{zhang2011sparse}, we recognize that attempting to estimate all $N$ covariance matrices $\bar{\vec{R}}_i$ using the data available at the BS would result in overfitting. Therefore, instead of estimating $N$ covariance matrices $\bar{\vec{R}}_i$, we estimate only $N_{\mathrm{c}}$ covariance matrices $\bar{\vec{R}}_{\mathcal{C}_l}$, $l=1,\ldots,N{\mathrm{c}}$. In particular, we make the assumption that all effective channels within a single cluster share the same spatial correlation structure\footnote{ It is worth noting that this assumption is not only technical but also highly justified in the context of MIMO channels, where closely located UEs are expected to have approximately the same covariance matrices \cite{studer2018channel}}, i.e., $\vec{x}_i \sim \mathcal{CN}(\vec{0},\bar{\gamma_i}\bar{\vec{R}}_{\mathcal{C}_l})$, ${\forall i\in \mathcal{C}_l}$, ${\forall l=1,\ldots,N_{\mathrm{c}}}$. Consequently,  the optimization problem \eqref{eq:Q_theta} with respect to $\bar{\vec{R}}_{\mathcal{C}_l}$   decoupled into $L$ subproblems as 
  \begin{equation}\label{eq:gamma_R_opt}
    \bar{\vec{R}}_{\mathcal{C}_l}^{(k)}= \displaystyle\underset{\bar{\vec{R}}_{\mathcal{C}_l} \succcurlyeq \vec{0}}{\max}~
-(L+Ld) \bar{\vec{R}}_{\mathcal{C}_l}-  \text{Tr} \bigg[\bar{\vec{R}}_{\mathcal{C}_l}^{-1}\bigg(\sum_{i \in \mathcal{C}_l}\frac{\vec{m}_i^{(k)}\vec{m}_i^{(k)\herm}+\Sigmab_i^{(k)}}{\bar{\gamma}_i}+L\vec{B}_{\mathcal{C}_l}\bigg)\bigg],  l=1,\ldots,N_{\mathrm{c}}.
\end{equation}
Therefore, by applying the first optimally condition, $\bar{\vec{R}}_{\mathcal{C}_l}$ is given as
\begin{equation}\label{eq:R_bar_sol}
    \bar{\vec{R}}_{\mathcal{C}_l}^{(k)}=\frac{1}{L+Ld}\bigg(\sum_{i \in \mathcal{C}_l}\frac{\vec{m}_i^{(k)}\vec{m}_i^{(k)\herm}+\Sigmab_i^{(k)}}{\bar{\gamma}_i}+L\vec{B}_{\mathcal{C}_l}\bigg), l=1,\ldots,N_{\mathrm{c}}. 
\end{equation}
Subsequently, we set $\bar{\vec{R}}_i^{(k)}=\bar{\vec{R}}_{\mathcal{C}_l}^{(k)}$, $\forall i \in \mathcal{C}_l$.
\subsection{Algorithm Implementation}
The details of the proposed  algorithm  termed EM-EP, are summarized in Algorithm 1.  EM-EP is run until $\fro{\vec{X}^{(k)}-\vec{X}^{(k-1)}}<\epsilon_{\mathrm{stp}}$  or until a maximum number of iterations  $k_{\mathrm{max}}$ is reached. 
Next, we outline two practical implementation considerations   of the EM-EP algorithm. 
\subsubsection{Non-positive Covariance Matrix}
While  $Q(\cdot)$ and $Q^{\mathrm{new}}(\cdot)$ have to be proper distributions in the EP framework, the approximated terms  $q_1(\cdot)$ and $q_2(\cdot)$, on the other hand, are not constrained to be proper distributions. In practice, some factors $q_2(\cdot)$ may become improper, resulting  in non-positive definite covariance matrices. In this work, for any non-positive definite matrix $\Sigmab_{2,i}$, $\forall i$, we add a small regularization parameter $\zeta>0$ to its diagonal elements to ensures that the covariance matrix is positive definite. Some  Alternative solutions have also been proposed \cite{vehtari2020expectation}.
\subsubsection{Pruning}
In practice, to overcome the computational complexity of EM-EP, at each iterations, we reduce the search space by ignoring  any  non-active UEs when updating the approximate factors. To this end, we use the posterior mean of the activity cluster indicator $\mathbb{E}_{f_{2,l}Q\ex{2,l}}[c_l]$ as a measure to prune non-active clusters. More precisely, for any cluster $l$ with $\mathbb{E}_{f_{2,l}Q\ex{2,l}}[c_l]<\epsilon_{\mathrm{thr}}$, all the effective channels are set to zero, i.e.,  $\vec{X}_{\mathcal{C}_l}=\vec{0}$, and they are pruned from the model.
\begin{algorithm}[t]
\DontPrintSemicolon
   \footnotesize \KwInput{ received signal $\vec{Y}$, pilot sequence matrix $\vec{\Phi}$, noise variance $\sigma^2$, $\epsilon$, $\epsilon_{\mathrm{stp}}$,  $\{\vec{B}_l\}_{l=1}^{N_\mathrm{c}}$,$k_{\mathrm{c}}$.}
\footnotesize \Kwinitialize{$\scriptsize \vec{m}_2^{(0)},\vec{\Sigmab}_2^{(0)},{k=1}$}
Compute the  $
\bar{\Sigmab}_1^{-1}  = \frac{1}{\sigma^2}\vecgreek{\Theta}\herm\vecgreek{\Theta}$ $\quad\quad\quad\vec{V}_1^{-1} \vec{\bar{m}}_1=\frac{1}{\sigma^2}\vecgreek{\Theta}\herm\vec{y} $

   \While{$k<k_\mathrm{max}$ $\mathrm{or}$ $\| \vec{X}^{(k)}-\vec{X}^{(k-1)} \|<\epsilon_{\mathrm{stp}}$ }
   {
 Compute the covariance $\hat{\Sigmab}_i\ex{2,l}$ and the mean $\hat{\vec{m}}_i\ex{2,l}$ of the cavity distribution  $\vec{m}_2$ using \eqref{moments_hat},  \;
 Compute the normalization term $G_{l,0}$ using \eqref{eq:G_0}\;
 
 Compute the  covariance $  \textrm{Var}_{f_{2,l}Q\ex{2,l}}[\vec{x}_i]$ and  the mean $\mathbb{E}_{f_{2,l}Q\ex{2,l}}[\vec{x}_i] $ using \eqref{q_new_moments}\;
 Compute the new  covariance $\hat{\Sigmab}_{2,i}$ and the mean $\hat{\vec{m}}_{2,i}$ of $q_2(\vec{X}_{\mathcal{C}_l},c_l)$ using \eqref{eq:Sigma2} and \eqref{eq:m_2}, respectively\;
  update the new posterior mean $\vec{m}_i$ and covariance $\Sigmab_i$ using \eqref{eq:Q_moments}  \;
 Update the set of hyper-parameters $\Xib$ using \eqref{eq:gamma_bar_sol} and \eqref{eq:R_bar_sol}
 \;

$\scriptsize k\leftarrow{k+1}$\;
}
\KwOutput{$\vec{X}=\vec{m}$}
\caption{EM-EP}
\end{algorithm}


\section{Alternative Solution via ADMM}\label{Sec:Cor-map}

Owing to the proposed \emph{the Hierarchical spike-and-slab} prior, and the efficient joint posterior approximation obtained using the EP framework, the proposed EM-EP algorithm  provides superior performance compared to the state-of-the-art algorithm, as we will show in the simulation analysis. However, the EM-EP algorithm, like most of EP-based methods, comes with the burden of high computational complexity. Therefore, in this section, we provide an alternative solution to the JUICE problem by relaxing the \emph{strong} spike-and-slab prior  by a weak prior based on the \emph{log-sum penalty} \cite{candes2008enhancing}, that still captures the essence of the \emph{ hierarchical structure} of the UEs activity pattern. Furthermore, we reformulate the JUICE as a MAP estimation that is solved via ADMM  in order to  offer a computationally-efficient  algorithm with closed-form updates that can be computed via simple analytical expressions. Next, we present in detail the proposed solution.
\vspace{-1.2cm}
\subsection{JUICE as MAP Estimation}\vspace{-.2cm}
In this section, we formulate the JUICE   as a MAP estimation problem in order to:  1) identify the active clusters, and  their corresponding active UEs, 2) estimate  the effective channel matrix $\vec{X}$, 3) estimate the unknown covariance matrices $\{\vec{R}_i\}_{i=1}^N$. Subsequently,
given the received signal $\vec{Y}$,  the MAP estimation problem  is expressed as

\begin{equation}
\begin{array}{ll}\label{eq:map_x_gamma}
\{\hat{\vec{X}},\hat{\gammab},\hat{\vec{R}}\}&\hspace{-3mm}=\underset{\vec{X},\gammab,\vec{R}}{\max}~\displaystyle p(\vec{X},\gammab,\vec{R}|\vec{Y})=\underset{\vec{X},\gammab,\vec{R}}{\max}~\displaystyle\frac{p(\gammab,\vec{R})p(\vec{X}|\gammab,\vec{R})p(\vec{Y}|\vec{X},\gammab,\vec{R})}{p(\vec{Y})}\\&\hspace{-3mm}\overset{(a)}{=}\underset{\vec{X},\gammab,\vec{R}}{\max}~\displaystyle{p(\gammab,\vec{R})p(\vec{X}|\gammab,\vec{R})p(\vec{Y}|\vec{X})}\\
&\hspace{-3mm}\overset{}{=}\underset{\vec{X},\gammab,\vec{R}}{\min}~\displaystyle{-\log\,p(\vec{Y}|\vec{X})}-\log\,p(\vec{X}|\gammab,\vec{R})-\log\,p(\gammab,\vec{R})\\
&\hspace{-3mm}\overset{(b)}{=}\underset{\vec{X},\gammab,\vec{R}}{\min}~\displaystyle\frac{1}{\sigma^2}\|\vec{Y}-\vec{\Phi}\vec{X}\tran\|_{\mathrm{F}}^{2}-\log\,p(\vec{X}|\gammab,\vec{R})-\log\,p(\vec{R})-\log\,p(\gammab) \end{array}
\end{equation}
where $(a)$ follows from the Markov chain ${\{\gammab,{\vec{R}}\}\rightarrow\vec{X}\rightarrow\vec{Y}}$ and  since the maximization is independent from $p(\vec{Y})$, and $(b)$ follows from the likelihood function of the received signal model  \eqref{eq::Y}.  Note that an alternative expression to the conditional PDF of $\vec{X}$ in \eqref{eq:P(X)_l} is given as
\begin{equation}\label{eq:P(x)weak}
       p(\vec{X}|\gammab,\vec{R})= \prod_{i=1}^Np(\vec{x}_i|\gamma_i,\vec{R}_i)=\prod_{i=1}^N\mathcal{CN}(\vec{x}_i;\vec{0},p_i\vec{R}_i)^{\mathbb{I}(\gamma_i=0)}, 
\end{equation}
where $\mathbb{I}(a)$ is an indicator function that takes the value 1 if $a\neq0$, and 0 otherwise. Note that $p(\vec{x}_i|\gamma_i,\vec{R}_i)$ in \eqref{eq:P(x)weak} implies that if $\gamma_i=0$, then $\vec{x}_i$ equals 0 with probability one, and in the case  of ${\gamma_i=1}$, the effective channel $\vec{x}_i$  follows a complex Gaussian distribution with zero mean  and covariance matrix $p_i\vec{R}_i$.

\vspace{-.4cm}\subsection{Cluster-Sparsity Promoting Prior via Log-Sum }\label{sec:p(gamma)}
Recall  the definition of the effective channel $\vec{x}_i=\sqrt{p_i}\gamma_i\vec{h}_i$, where the true indicator variable $\gamma_i\in \{0,1\}$ controls the sparsity of $i$th UE. Thus, assigning a sparsity-promoting prior to $p(\gammab)$ is the key to obtain a  sparse solution to \eqref{eq:map_x_gamma}. A conventional  choice for a \emph{tractable} sparsity-promoting prior $p(\gammab)$ is the \emph{log-sum} penalty prior ${\sum_{i=1}^{N}\log(\gamma_i+\epsilon_0)}$ as it  resembles most closely the canonical $\ell_0$-norm  when ${\epsilon_{0} \rightarrow 0}$. Subsequently, we define the following  sparsity prior $p(\gammab)$  
\begin{equation}\label{eq:T1}
p(\gammab)\propto J_{\mathrm{s}}(\gammab)=\sum_{i=1}^{N}\log(\gamma_i+\epsilon_0). 
\end{equation}
Although the prior \eqref{eq:T1} is an appropriate choice as it: 1) promotes sparsity, 2) is separable across the UEs, it ignores the hierarchical structure of the UEs' activity pattern. Therefore, to account for the cluster-level sparsity, we propose the  \emph{cluster-sparsity-promoting prior} $J_c(\cdot)$ that correlates the UE activity indicators variables  belonging to the same cluster, i.e., $\gamma_i$, ${i\in \mathcal{C}_l}$ with the same log-sum penalty. More precisely, we propose the following prior function:
\begin{equation}\label{eq:T2}
p(\gammab) \propto J_{\mathrm{c}}(\gammab)= \sum_{l=1}^ {N_{\mathrm{c}}}\log\Big(\sum_{i \in \mathcal{C}_l}\gamma_i+\epsilon_0\Big). 
\end{equation}
Note that $J_{\mathrm{c}}(\gammab)$ promotes quite stringently solutions that have clustered sparsity as it has the tendency to enforce all UEs within each cluster to be detected active even if only one UE is active, being thereby susceptible to high false alarm error rate. Thus, $J_{\mathrm{c}}(\gammab)$ would face  robustness issues in the instances where the UEs activity pattern does not exhibit a clustered structure. 

\vspace{-.4cm}\subsection{Proposed ADMM Solution}
Before we derive the ADMM based solution for \eqref{eq:map_x_gamma}, we make two technical choices:
    1) the binary nature of $\gammab$ renders the objective function in \eqref{eq:map_x_gamma} intractable for large $N$. Thus, to overcome this challenge, we note that finding the index set $\{i \mid \gamma_i \neq0,\;i\in\mathcal{N}\}$ is equivalent to finding the index set $\{i \mid \|\vec{x}_i\| >0,\;i\in\mathcal{N}\}$. Thus, we can eliminate the variable  $\gammab$ from the MAP problem by approximating each $\gamma_i$ by $\|\vec{x}_i\|$ and by relaxing $p(\gammab)$ by an equivalent prior function $p(\vec{X})$ which depends on  $\|\vec{x}_i\|$, $\forall i \in \mathcal{N}$. 
    2) Similarly to Section \ref{R_update_EP}, we resort to estimating a shared covariance matrix for all the UEs within the same cluster.
    
By using the aforementioned arguments and introducing the regularization weights $\beta_1$, $\beta_2$,  and $\beta_3$ that control the emphasis on the priors with respect to the measurement fidelity term,  the MAP estimation problem \eqref{eq:map_x_gamma} can be equivalently  rewritten as
\begin{equation}\label{map_S}
\begin{array}{ll}
       \!\!\{\hat{\vec{X}},\hat{\vec{R}}\}\!=\!\underset{\vec{X},\vec{R}}{\min}~\displaystyle\frac{1}{2}\|\vec{Y}-\vec{\Phi}\vec{X}\tran\|_{\mathrm{F}}^{2} -\beta_1\log p(\vec{X})+\beta_2\sum_{l=1}^{N_{\mathrm{c}}}\sum_{i \in \mathcal{C}_l}\frac{\vec{x}_{i}\herm\vec{R}_{\mathcal{C}_l}\vec{x}_{i}}{p_i} \\\displaystyle\!\!-\beta_2\sum_{l=1}^{N_{\mathrm{c}}}  \log |\vec{R}_{\mathcal{C}_l}| \sum_{i \in \mathcal{C}_l}p_i^M \|\vec{x}_i\| \!-\! \beta_3 L\sum_{l=1}^{N_{\mathrm{c}}} \big( d \log |\vec{R}_{\mathcal{C}_l}|\displaystyle \!+\!  \mathrm{tr}(\vec{B}_l^{-1}\vec{R}_{\mathcal{C}_l})\big).
\end{array}
\end{equation}

Now, we propose  an iterative solution based on a hierarchical algorithm with two loops, an \emph{outer loop}  and an  \emph{inner loop}. The central idea is to alternate $p(\vec{X})$ over $J_{\mathrm{c}}(\cdot)$ \eqref{eq:T2} and $J_{\mathrm{s}}(\cdot)$ \eqref{eq:T1}. More precisely, in the outer loop, the algorithm enforces the detection of active clusters via the cluster-level sparsity-promoting function $J_{\mathrm{c}}$. Subsequently, the algorithm runs an inner loop over the just-estimated active clusters to detect the individual active UEs belonging to them by using the sparsity-promoting prior $J_{\mathrm{s}}$. The algorithm details are presented next.

\subsubsection{Outer Loop}
As we aims to detect the set of the active clusters first,  we enforce $p(\vec{X})$ to promote the cluster-level sparsity by $- \log p(\vec{X})=\sum_{l=1}^{N_{\mathrm{c}}}\log(\sum_{i \in \mathcal{C}_l}\|\vec{x}_i\|+\epsilon_0)$. Since $-\log p(\vec{X})$ is concave, we apply a majorization-minimization (MM) approximation \cite{sun2016majorization} to linearize $- \log p(\vec{X})\approx \sum_{i=1}^N q_i^{(k_{\mathrm{c}})}\|\vec{x}_i\|$, 
where $q_i^{(k_{\mathrm{c}})}=\big(\sum_{i \in \mathcal{C}_l}\|\vec{x}_i^{(k_{\mathrm{c}})}\|+\epsilon_0\big)^{-1}$ and  $k_{\mathrm{c}}$ is the MM iteration index  for the outer loop. 
Thus,  the relaxed version of the  problem \eqref{map_S} can be solved iteratively as  
\begin{equation}
\begin{array}{ll}
      &\{\hat{\vec{X}}^{(k_{\mathrm{c}}+1)},\vec{R}^{(k_{\mathrm{c}}+1)}\} =\displaystyle\underset{\vec{X},\vec{R}}{\min}~\displaystyle\frac{1}{2}\|\vec{Y}-\vec{\Phi}\vec{X}\tran\|_{\mathrm{F}}^{2}+\beta_1 \sum_{i=1}^N q_i^{(k_{\mathrm{c}})} \|\vec{x}_i\| \displaystyle\\&+\beta_2\sum_{l=1}^{N_{\mathrm{c}}}\sum_{i \in \mathcal{C}_l}\vec{x}_{i}\herm\vec{R}_{\mathcal{C}_l}\vec{x}_{i}+\sum_{l=1}^{N_{\mathrm{c}}} \mu^{(k_{\mathrm{c}})} \log |\vec{R}_{\mathcal{C}_l}| +\beta_3 L\sum_{l=1}^{N_{\mathrm{c}}} \mathrm{tr}(\vec{B}_l\vec{R}_{\mathcal{C}_l}^{-1}),
\end{array}\label{map_K_c}
\end{equation}
where ${\scriptsize \mu^{(k_{\mathrm{c}})}=\big(\beta_2  \displaystyle\sum_{i \in \mathcal{C}_l} p_i^M q_i^{(k_{\mathrm{c}})}\|\vec{x}_i^{(k_{\mathrm{c}})}\| + \beta_3 L d\big)}$.
 
We develop a computationally efficient ADMM solution for \eqref{map_K_c} through a set of sequential update rules, each computed in a closed-form. To this end, we introduce two splitting variables ${\vec{Z},\vec{V}\in \mathbb{C}^{M\times N}}$ and the Lagrange dual variable  $\Lambdab_v,\Lambdab_z$ and define the set of variables to be estimated as $\Omega=\{\vec{X},\vec{R},\vec{Z},\vec{V},\Lambdab_{\mathrm{z}},\Lambdab_{\mathrm{v}}\}$. Subsequently, we write the augmented Lagrangian as
\vspace{-.3cm}\begin{equation}\label{eq:lagrange}
  \begin{array}{ll}
       &\mathcal{L}(\Omega)=\displaystyle\frac{1}{2}\|\vec{Y}-\vec{\Phi}\vec{Z}\tran\|_{\mathrm{F}}^{2} +\beta_1 \sum_{i=1}^N q_i^{(k_{\mathrm{c}})} \|\vec{x}_i\|+\beta_2\sum_{i=1}^{N}\vec{v}_{i}\herm\vec{R}_{\mathcal{C}_l}\vec{v}_{i}+\frac{\rho}{2}\|\vec{X}-\vec{V}+\frac{1}{\rho}\vecgreek{\Lambda}_{\mathrm{v}}\|_{\mathrm{F}}^2\\\displaystyle &+\displaystyle\sum_{l=1}^{N_{\mathrm{c}}}\big( \mu^{(k_{\mathrm{c}})} \log |\vec{R}_{\mathcal{C}_l}|  +\beta_3 L \mathrm{tr}(\vec{B}_l^{-1}\vec{R}_{\mathcal{C}_l})\big)+\displaystyle\frac{\rho}{2}\|\vec{X}-\vec{Z}+\displaystyle\frac{1}{\rho}\vecgreek{\Lambda}_{\mathrm{z}}\|_{\mathrm{F}}^2 -\displaystyle\frac{\fro{\vecgreek{\Lambda}_{\mathrm{z}}}}{2\rho}
     -\displaystyle\frac{\fro{\vecgreek{\Lambda}_{\mathrm{v}}}}{2\rho}.
\end{array}  
\end{equation}
The ADMM solves to the optimization problem \eqref{map_K_c}  by minimizing the augmented Lagrangian  $\mathcal{L}(\Omega)$ in \eqref{eq:lagrange}  over the primal variables $(\vec{Z},\vec{V},\vec{X},\Sigmab)$, followed by updating the dual variables $(\vecgreek{\Lambda}_{\mathrm{z}},\vecgreek{\Lambda}_{\mathrm{v}})$ \cite{boyd2011distributed}. The primal variable update is given as
\begin{spacing}{1.3}
\begin{equation}\label{ADMM_subproblems}  
\begin{cases}
         \vec{Z}^{(k_{\mathrm{c}}+1)}=
   \displaystyle\min_{\vec{Z}}\frac{1}{2}\| \vec{\Phi}\vec{Z}\tran-\vec{Y}\|_{\mathrm{F}}^2+ \frac{\rho}{2} \Vert\vec{X}^{(k_{\mathrm{c}})} -\vec{Z} +\frac{1}{\rho}\vecgreek{\Lambda}_{\mathrm{z}}^{(k_{\mathrm{c}})} \|_{\mathrm{F}}^2\\
      \vec{V}^{(k_{\mathrm{c}}+1)}=\displaystyle\min_{\vec{V}}\beta_2  \sum_{i=1}^{N}  \vec{v}_i\herm\vec{R}_{\mathcal{C}_l}^{(k_{\mathrm{c}})}\vec{v}_i  +\frac{\rho}{2} \Vert   \vec{X}^{(k_{\mathrm{c}})} - \vec{V} +\frac{\vecgreek{\Lambda}_{\mathrm{v}}^{(k_{\mathrm{c}})}}{\rho} \|_{\mathrm{F}}^2, \\
    \vec{X}^{(k_{\mathrm{c}}+1)}=\displaystyle\min_{\vec{X}}  \sum_{i=1}^{N} \alpha_i^{(k_{\mathrm{c}})} \Vert  \vec{x}_i\|+ \frac{\rho}{2} \|\vec{X}-\vec{C}^{(k_{\mathrm{c}}+1)} \|_{\mathrm{F}}^2
,\\
     \vec{R}_{\mathcal{C}_l}^{(k_{\mathrm{c}}+1)}\!\!=\!\displaystyle\min_{\vec{R}_{\mathcal{C}_l}}  \beta_2\sum_{i \in \mathcal{C}_l}\!\vec{v}_{i}^{(k_{\mathrm{c}}+1)\herm}\!\vec{R}_{\mathcal{C}_l}\vec{v}_{i}^{(k_{\mathrm{c}}+1)}\!+\!\mu^{(k_{\mathrm{c}}+1)}\! \log |\vec{R}_{\mathcal{C}_l}|+\beta_3 L \mathrm{tr}(\vec{B}_l^{-1}\vec{R}_{\mathcal{C}_l}),~ \forall l,
    \end{cases}       
\end{equation}
\end{spacing}

where   ${\vec{C}^{(k_{\mathrm{c}})}=\dfrac{1}{2}\big( \vec{Z}^{(k_{\mathrm{c}}+1)}+\vec{V}^{(k_{\mathrm{c}}+1)}-\displaystyle\dfrac{\vecgreek{\Lambda}_{\mathrm{z}}^{(k_{\mathrm{c}})}+\vecgreek{\Lambda}_{\mathrm{v}}^{(k_{\mathrm{c}})}}{\rho}\big)}$ and 
${\alpha_i^{(k_{\mathrm{c}})}= \big(\beta_1 q_i^{(k_{\mathrm{c}})}-\beta_2 \log|\vec{R}_{\mathcal{C}_l}^{(k_{\mathrm{c}})}|q_i^{(k_{\mathrm{c}})}\big)}$.

\subsubsection{Inner Loop}
After running the outer loop for some pre-defined $K_{\mathrm{c}}$ iterations, we detect the set of the estimated active clusters $\mathcal{\hat{S}}=\bigcup_{j \in \mathcal{J}} \mathcal{C}_j$, where $l \in \mathcal{J}$ if there exists $i \in \mathcal{C}_l$ such that $\|\vec{x}_i\|>\epsilon_{\mathrm{thr}}$,
where $\epsilon_{\mathrm{thr}}>0$ is a small predefined parameter.
In the inner loop, the proposed algorithm aims to detect  the  active UEs belonging to $\displaystyle\mathcal{\hat{S}}$  by using the separable sparsity-promoting prior
${-\log  p(\vec{X}) \propto  \sum_{i \in \mathcal{\hat{S}}}\log(\|\vec{x}_i\|+\epsilon_0)}$. Furthermore, we apply the MM approximation to linearize the concave   $-\log  p(\vec{X})\approx \sum_{i \in \mathcal{\hat{S}}} g_i^{(k_{\mathrm{c}})}\|\vec{x}_i\|,$
where $k_\mathrm{u}$ is inner loop iteration index and 
${g_i^{(k_\mathrm{u})}=\big( \|\vec{x}_i^{(k_\mathrm{u})}\|+\epsilon_0\big)^{-1}}$. Subsequently, the optimization problem for the inner loop is given by
\begin{equation}
\begin{array}{ll}
&\displaystyle\hspace{-4mm}\{\hat{\vec{X}}_\mathcal{\hat{S}}^{(k_\mathrm{u}+1)},\vec{R}_{\mathcal{J}}^{(k_\mathrm{u}+1)}\} =\displaystyle\underset{\vec{X}_\mathcal{\hat{S}},\vec{R}}{\min}~\displaystyle\frac{1}{2}\|\vec{Y}-\vec{\Phi}_\mathcal{\hat{S}}\vec{X}_\mathcal{\hat{S}}\tran\|_{\mathrm{F}}^{2}+\beta_1 \sum_{i \in \mathcal{\hat{S}}} g_i^{(k_\mathrm{u})} \|\vec{x}_i\| \displaystyle+\beta_2\sum_{l\in \mathcal{J}}\sum_{i \in \mathcal{\hat{S}}_l}\vec{x}_{i}\herm\vec{R}_{\mathcal{C}_l}\vec{x}_{i}\\
      &+\beta_2\sum_{l\in \mathcal{J}}  \log |\vec{R}_{\mathcal{C}_l}| \big(\sum_{i \in \mathcal{C}_l} g_i^{(k_\mathrm{u})}\|\vec{x}_i\|+\beta_3Ld\big)  +\beta_3 L\sum_{l\in \mathcal{J}} \mathrm{tr}(\vec{B}_l^{-1}\vec{R}_{\mathcal{C}_l}),
\end{array}\label{map_S_inner}
\end{equation}
where $\vec{\Phi}_\mathcal{\hat{S}}$ and $\vec{X}_\mathcal{\hat{S}}$ denote the matrices $\vecgreek{\Phi}$ and $\vec{X}$, respectively, restricted to the set $\mathcal{\hat{S}}$.


\vspace{-.2cm}
\subsubsection{ADMM Update Rules}
The details of the proposed algorithm in section \ref{Sec:Cor-map}, termed corr-MAP-ADMM, is summarized in   Algorithm 2. Note that
owing to the  \emph{ proposed splitting techniques} in \eqref{eq:lagrange}, all the sub-problems in \eqref{ADMM_subproblems} are convex, thus, can be solved analytically via \emph{ closed-form formulas}. Further, the optimization over $\vec{X}$, $\vec{V}$, and $\Sigmab$ is separable over the UEs and the clusters, allowing for \emph{parallel} updates. Thus, the exact solution to \eqref{ADMM_subproblems} is given in steps 3-6 in Algorithm 2, and we follow the same analogy for to solve \eqref{map_S_inner} in the inner loop.

%
\begin{spacing}{1.5}
    \begin{algorithm}[t]
\DontPrintSemicolon
   \footnotesize \KwInput{$\vec{\Phi}$, $\{\vec{B}_l\}_{l=1}^C$,
   $\scriptsize\beta_1$,$\beta_2$, $\beta_3$ ,$\rho$, $\epsilon_0$, $\epsilon$,   $k_{u_{\mathrm{max}}}$, $k_{c_{\mathrm{max}}}$, $\scriptsize K_{\mathrm{c}}$.}
\footnotesize \Kwinitialize{$\scriptsize \vec{X}^{(0)},\vec{V}^{(0)},\vec{Z}^{(0)},\vecgreek{\Lambda}_{\mathrm{v}}^{(0)},\vecgreek{\Lambda}_{\mathrm{z}}^{(0)}, {k_\mathrm{u}=1, k_{\mathrm{c}}=1.}$}
\footnotesize BS receives $\vec{Y}$, compute and store $\big( \vec{\Phi}\tran \vec{\Phi}^*+\rho \vec{I}_N\big)^{-1}$ 
 
   \While{$k_{\mathrm{c}}<k_{c_{\mathrm{max}}}$ $\mathrm{or}$ $\| \vec{X}^{(k_{\mathrm{c}})}-\vec{X}^{(k-1)} \|<\epsilon$ }
   {

  $\tiny\vec{Z}^{(k_{\mathrm{c}}+1)}=  (\rho \vec{X}^{(k_{\mathrm{c}})} +\vec{\Lambda}^{(k_{\mathrm{c}})}+\vec{Y}\tran\vec{\Phi}^*)( \vec{\Phi}\tran \vec{\Phi}^*+\rho \vec{I}_N)$ 
 \; 
  $\scriptsize\vec{v}_i^{(k_{\mathrm{c}}+1)}=  \big(\beta_2\Sigmab_i^{(k_{\mathrm{c}})}+\rho\vec{I}_M \big)^{-1}(\rho \vec{x}_i^{(k)}+\vecgreek{\lambda}_{\mathrm{v}i}^{(k_{\mathrm{c}})}), {i=1,\ldots,N}$ 
 \;
 $\scriptsize\vec{x}_i^{(k_{\mathrm{c}}+1)}= \frac{\max{\big\{0,\norm{\vec{c}_i^{(k_{\mathrm{c}})}}-\frac{\alpha_i^{(k_{\mathrm{c}})}}{2\rho}\big\}}\vec{c}_i^{(k_{\mathrm{c}})}}{ \norm{\vec{c}_i^{(k_{\mathrm{c}})}}}, {i=1,\ldots,N}$
 \;
  $ \scriptsize\vec{R}_{\mathcal{C}_l}^{(k_{\mathrm{c}})+1}=\frac{1}{\mu^{(k_{\mathrm{c}})}}\big(\beta_2\sum_{i \in \mathcal{C}_l}\vec{v}_{i}\vec{v}_{i}\herm+\beta_3L\vec{B}_l\big),l=1,\ldots,C$
 \;
 $\scriptsize\vecgreek{\Lambda}_{\mathrm{z}}^{(k_{\mathrm{c}}+1)}=  \vecgreek{\Lambda}_{\mathrm{z}}^{(k_{\mathrm{c}})}+\rho\big(   \vec{X}^{(k_{\mathrm{c}}+1)}-\vec{Z}^{(k_{\mathrm{c}}+1)} \big)$
 \;
 $\scriptsize\vecgreek{\Lambda}_{\mathrm{v}}^{(k_{\mathrm{c}}+1)}=  \vecgreek{\Lambda}_{\mathrm{v}}^{(k_{\mathrm{c}})}+\rho\big(   \vec{X}^{(k_{\mathrm{c}}+1)}-\vec{V}^{(k_{\mathrm{c}}+1)} \big)$
 \;
\uIf{\big($k_{\mathrm{c}}\ \mathrm{mod}\ K_{\mathrm{c}}\big) = 0$}{
 $\scriptstyle\mathcal{\hat{S}}=\bigcup_{j \in \mathcal{J}} \mathcal{C}_j$, $\{l \in \mathcal{J}: \exists i \in \mathcal{C}_l| \|\vec{x}_i\|>\epsilon\}$
 \;
  \While{\scriptsize$k_\mathrm{u}<k_{\mathrm{u}_{{\mathrm{max}}}}$ }
   {
 Solve \eqref{map_S_inner} using the similar update rules as step 3-6, but using $\scriptsize g_i^{(k_\mathrm{u})}=\big(\|\vec{x}_i^{(k_\mathrm{u})}+\epsilon_0\|\big)^{-1}$\;
$k_{\mathrm{u}}\leftarrow{k_{\mathrm{u}}+1}$\;
} 
$\scriptstyle\vec{X}_{\mathcal{\hat{S}}}^{(k_{\mathrm{c}})}=\vec{X}_{\mathcal{\hat{S}}}^{(k_\mathrm{u})}$, $\scriptstyle\vec{Z}_{\mathcal{\hat{S}}}^{(k_{\mathrm{c}})}=\vec{Z}_{\mathcal{\hat{S}}}^{(k_\mathrm{u})}$, $\scriptstyle\vec{V}_{\mathcal{\hat{S}}}^{(k_{\mathrm{c}})}=\vec{V}_{\mathcal{\hat{S}}}^{(k_\mathrm{u})}$, $\scriptstyle\vecgreek{\Lambda}_{\mathrm{z}_{\mathcal{\hat{S}}}}^{(k_{\mathrm{c}})}=\vecgreek{\Lambda}_{\mathrm{z}_{\mathcal{\hat{S}}}}^{(k_{\mathrm{u}})}$,  $\scriptstyle\vecgreek{\Lambda}_{\mathrm{v}_{\mathcal{\hat{S}}}}^{(k_{\mathrm{c}})}=\vecgreek{\Lambda}_{\mathrm{v}_{\mathcal{\hat{S}}}}^{(k_{\mathrm{u}})}$\;
  $k_\mathrm{u}=1$,\;
 }
$\scriptsize k_{\mathrm{c}}\leftarrow{k_{\mathrm{c}}+1}$\;
}
\caption{\it{corr}-MAP-ADMM}
\end{algorithm}
\end{spacing}

\vspace{-.2cm}

\section{Simulation Result}
This section quantifies the  performance and the robustness of the proposed  algorithms and compare them to existing sparse recovery  algorithms in terms of active UEs identification accuracy,  channel estimation quality, and convergence behaviour. 
\vspace{-.5cm}\subsection{Simulation Setup}
We consider a network with a single BS serving a  ${N=200}$ UEs distributed equally over ${N_{\mathrm{c}}=20}$ clusters, with a total of active $K=16$.   Each UE is assigned a  unique unit-norm pilot sequence drawn  from an i.i.d. complex Bernoulli distribution. 
We set each $\vec{B}_l,  \forall l $, as $\vec{B}_l=\zeta \Psib_l+(1-\zeta) \frac{1}{L}\sum_{i\in \mathcal{C}_l}\vec{R}_i$, where $\Psib_l$ is a random positive-definite Hermitian matrix to model the error in the prior knowledge on the  covariance matrix $\vec{R}_{\mathcal{C}_l}$, whereas the parameter $\zeta$ controls the level of average  mismatch between $\vec{R}_i$ and $\vec{B}_l$, $\forall i \in \mathcal{C}_l$, and it is set as $\zeta=0.1$.  
\subsection{Metrics and Baselines}\label{sec_results_baselines}
The  performance is evaluated in terms of: 1) channel estimation quality measured by the normalized mean
square error (NMSE) given as $\frac{\mathbb{E}\left [\| \vec{X}-\hat{\vec{X}}_\mathcal{S}\|_{\mathrm{F}}^2 \right ]}{\mathbb{E}\left[\| \vec{X}\|_{\mathrm{F}}^2 \right]}$, 
2) activity detection accuracy  using the support recovery  rate (SRR), defined as $\frac{\vert \mathcal{S} \cap \hat{\mathcal{S}}\vert}{\vert \mathcal{S} - \hat{\mathcal{S}}\vert+K}$, where $\hat{\mathcal{S}}$ denotes the detected support by a given algorithm. 
We compare  the performances of the proposed algorithms against three algorithms that solve MMV sparse recovery problem, namely, 
 1) iterative reweighted $\ell_{2,1}$ (IRW-$\ell_{2,1}$) via ADMM  \cite{Djelouat-Leinonen-Juntti-21-icassp}, 2) MAP-ADMM \cite{djelouat2021spatial}, and  3) T-SBL \cite{zhang2011sparse}, where for  MAP-ADMM and T-SBL,  both the second-order statistics and the noise variance are known at the BS.  Furthermore,  For an optimal NMSE benchmark, we consider the oracle minimum mean square error (MMSE) estimator that is provided with an ``oracle'' knowledge on the true set of active UEs. 

In practice, the regularization parameters of  the sparse recovery algorithms needs to be fine-tuned. For a fair comparison, all the parameters have been empirically tuned via cross-validation in advance and then fixed such that they provide their best performance.

\vspace{-.5cm}\subsection{Correlated Activity Pattern}
First, we investigate the performance of the proposed EM-EP and corr-MAP-ADMM and compare it to the baseline algorithms. To this end,  the  16  active UEs are distributed over 2 active clusters, each with 8 active UEs.

\subsubsection{Effect of Pilot Length}
First, we quantify the  performance of the proposed algorithms in compassion to MAP-ADMM, T-SBL and IRW-$\ell_{2,1}$ for different pilot sequence lengths with fixed $\text{SNR}=16$~dB. 

\begin{figure}[t!]
\centering
   \begin{subfigure}[]{0.48\textwidth}    \centering\includegraphics[scale=.4]{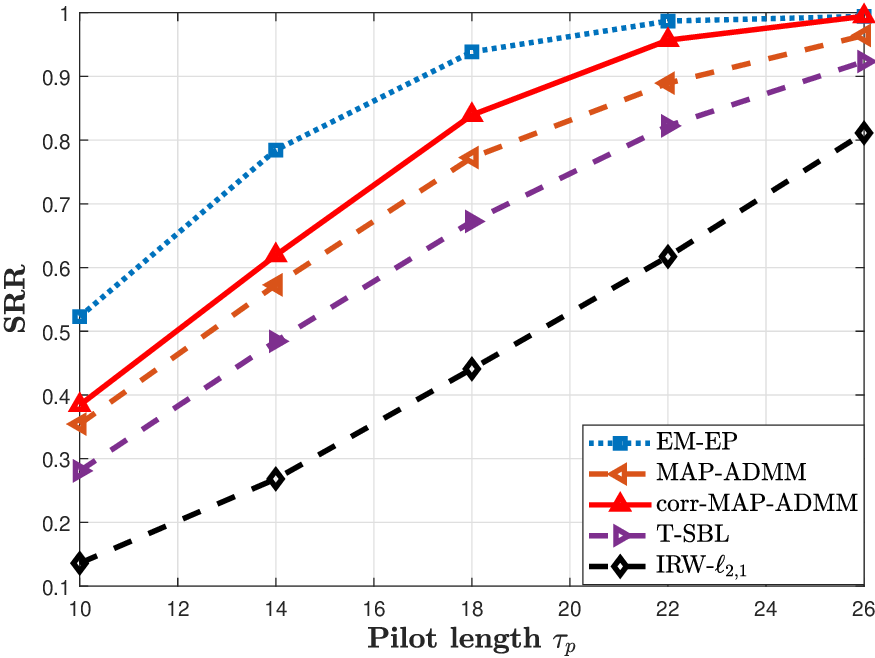}  \caption{}
    \label{mse_clus}
\end{subfigure}
   \hfill
    \begin{subfigure}[]{0.48\textwidth}
  \centering\includegraphics[scale=.4]{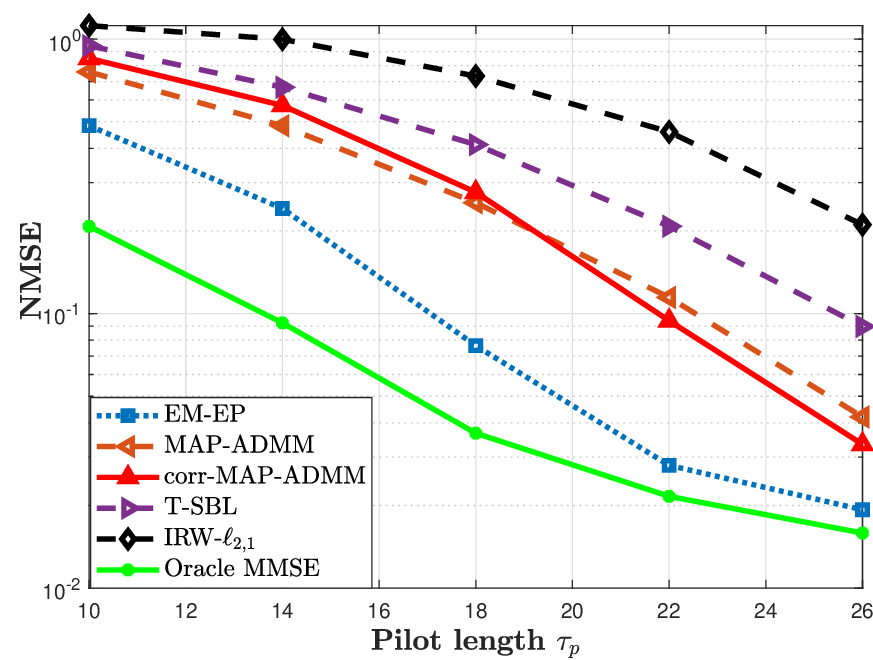}   \caption{}
     \label{mse_clus}
\end{subfigure}\vspace{-3mm}
\caption{JUICE Performance for the different algorithms against pilot sequence length $\tau_{\mathrm{p}}$ in terms of: (a) SRR, (b) NMSE. }
\label{fig:results_cluster}
\vspace{-6mm}
\end{figure}

Fig.\ \ref{fig:results_cluster}(a) depicts the SRR performance, where  the results show clearly that the proposed EM-EP and corr-MAP-ADMM algorithms provide the best activity detection accuracy as they achieve the highest SSR compared to all other algorithms. For instance, EM-EP achieves the same SRR as MAP-ADMM and T-SBL using  $25~\%$ shorter pilot sequence length. 
Furthermore, the results show that the EM-EP provides a comparable gain over corr-MAP-ADMM  in the low pilot length regime. Fig.\ \ref{fig:results_cluster}(b) depicts the channel estimation performance for the different recovery algorithms in terms of NMSE. Again, the obtained results highlight the remarkable gains in channel estimation accuracy obtained by using the EM-EP, which significantly outperforms all other algorithms. Indeed, even with a 30\% reduction in pilot length used, EM-EP achieves the same performance as corr-MAP-ADMM, MAP-ADMM and T-SBL.  Moreover, the proposed corr-MAP-ADMM moderately outperforms T-SBL and provides almost similar performance compared to MAP-ADMM. These results can be explained by the fact that the EM-EP, using the hierarchical spike-and-slab prior, efficiently captures the intrinsic structure of the sparsity model to improve the solution in terms of recovery rate and channel estimation. 

\subsubsection{Effect of SNR}
\begin{figure}[t!]\centering
 \begin{subfigure}[b]{0.48\textwidth}    \centering\includegraphics[scale=.4]{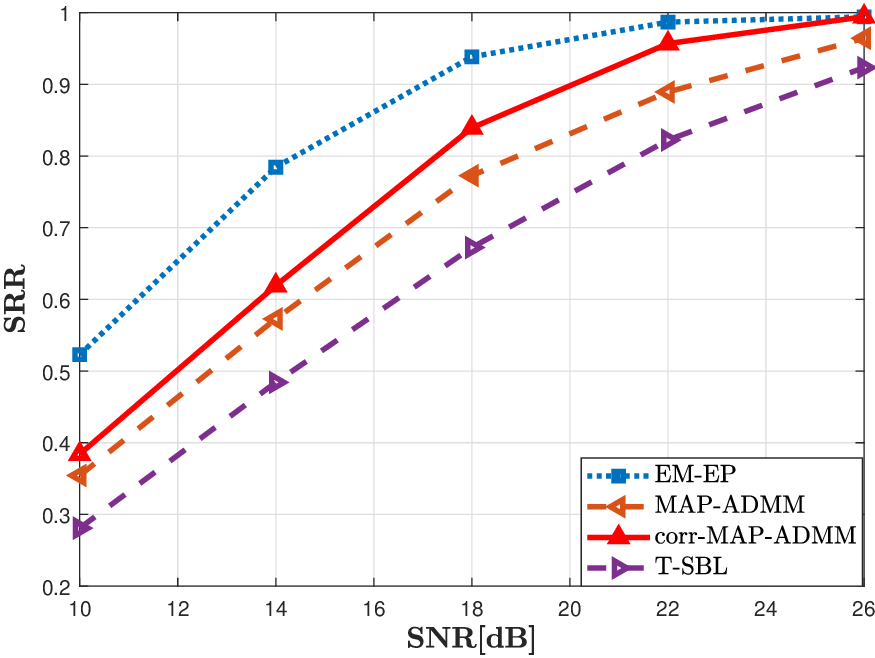}  \caption{}
    \label{mse_clus}
\end{subfigure}
   \hfill
    \begin{subfigure}[b]{0.48\textwidth}
\centering  \includegraphics[scale=.4]{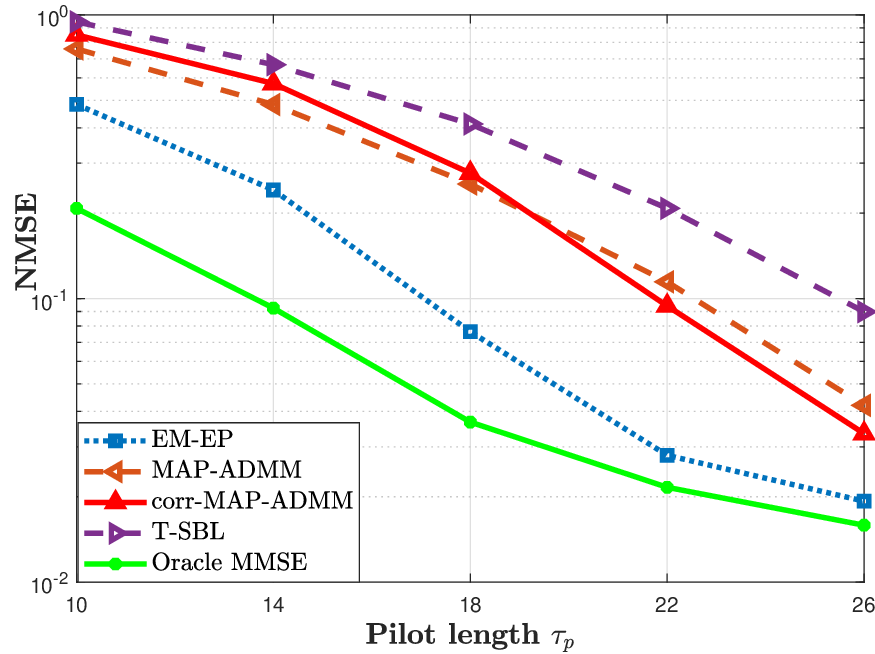}   \caption{}
     \label{mse_clus}
\end{subfigure}
\caption{JUICE Performance for the different algorithms against SNR in terms of: (a) SRR, (b) NMSE.  }
\label{fig:results_snr}
\end{figure}

Now we fix the pilot sequence length at $\tau_{\mathrm{p}}=24$, and examine the impact of SNR on the  performance of different algorithms.  Fig.\  \ref{fig:results_snr}(a) illustrates the SRR for different algorithms, where  it is evident that the proposed EM-EP and corr-MAP-ADMM provide the best activity detection accuracy, demonstrating a remarkable SRR gain in the lower SNR range. Additionally,  Fig.\  \ref{fig:results_snr}(b) depicts the channel estimation quality in terms of NMSE and shows that  corr-MAP-ADMM provides roughly the same performance as MAP-ADMM and T-SBL. Furthermore, the results emphasize the superiority of the proposed EM-EP in terms of channel estimation quality, providing an approximate gain of $4~$dB compared to MAP-ADMM and T-SBL.
\subsubsection{Effect of the Number of BS Antennas}
\begin{figure*}[t]
    \centering
    \includegraphics[scale=.37]{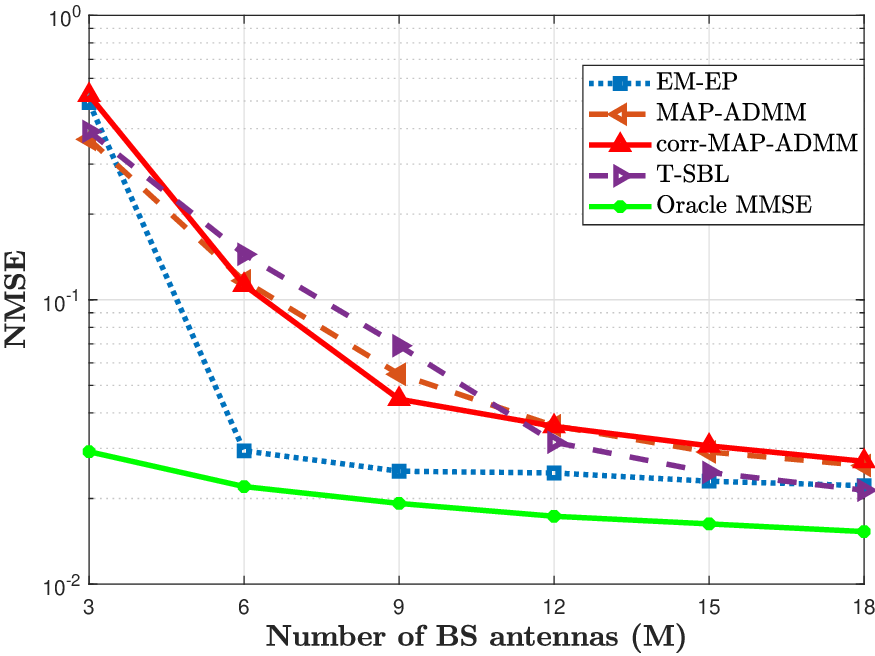}
    \label{fig:SRR_M}
\caption{JUICE performance in terms of  NASE versus the  number of BS antennas $M$ for $N=200$,   $\tau_{\mathrm{p}}=20$, and  ${\text{SNR}=15}$~dB.}
\label{fig:M}
\end{figure*}

\vspace{-4mm}
Fig. \ref{fig:M} illustrates the effect of the number of BS antennas ($M$) on the  performance of the proposed algorithms in terms of NMSE. As expected, all the algorithms experience an improvement in  the channel estimation quality  as $M$ increases. Notably,  EM-EP algorithm substantially outperforms  the other sparse recovery algorithms, with a remarkable gains  in the lower range for BS antennas, i.e.,  ${M<12}$. In addition, corr-MAP-ADMM provide the second best performance for  ${M<12}$.  However, it is important to note that as $M$ increases, the performance gap between all algorithms gradually decreases. This can be attributed to the fact that having more measurements from additional antennas would yields more gains compared to incorporating any side information, such as correlated activity structure.

In summary, the results in Fig. \ref{fig:results_cluster}, Fig. \ref{fig:results_snr}, and Fig. \ref{fig:M} clearly demonstrate the substantial gains achieved by 
incorporating the structured activity pattern when designing  solutions for the JUICE problem. Indeed, even with no prior knowledge on the CDI,  the proposed algorithms outperform   state-of-the-art recovery algorithms that utilizes  prior knowledge on the full CDI.
\vspace{-.5cm}\subsection{Robustness to Model Mismatch}
\begin{figure}[t!]\centering 
   \begin{subfigure}[b]{0.48\textwidth}\centering    \centering\includegraphics[scale=.4]{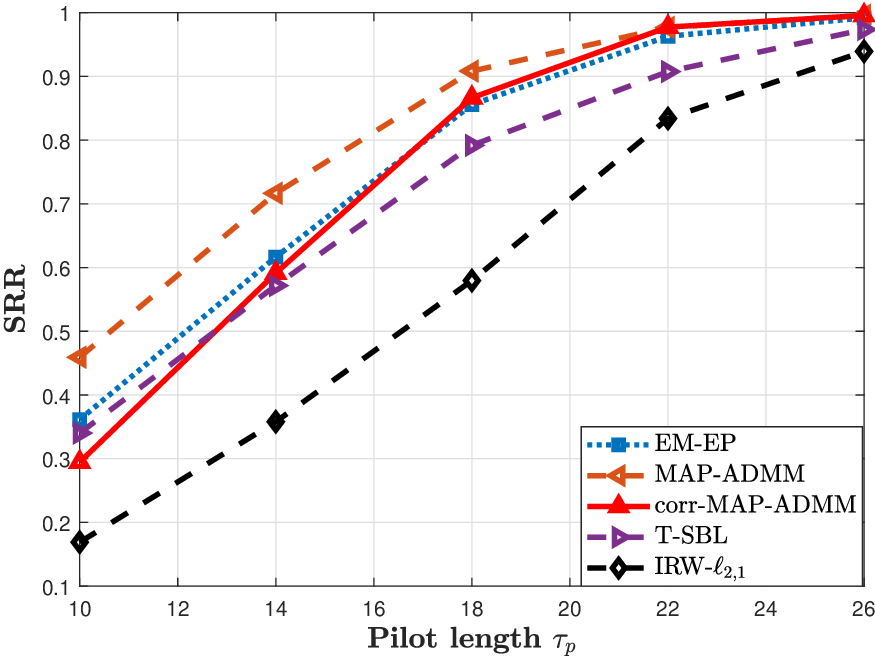}  \caption{}
    \label{mse_clus}
\end{subfigure}
   \hfill
    \begin{subfigure}[b]{0.48\textwidth}\centering 
  \includegraphics[scale=.4]{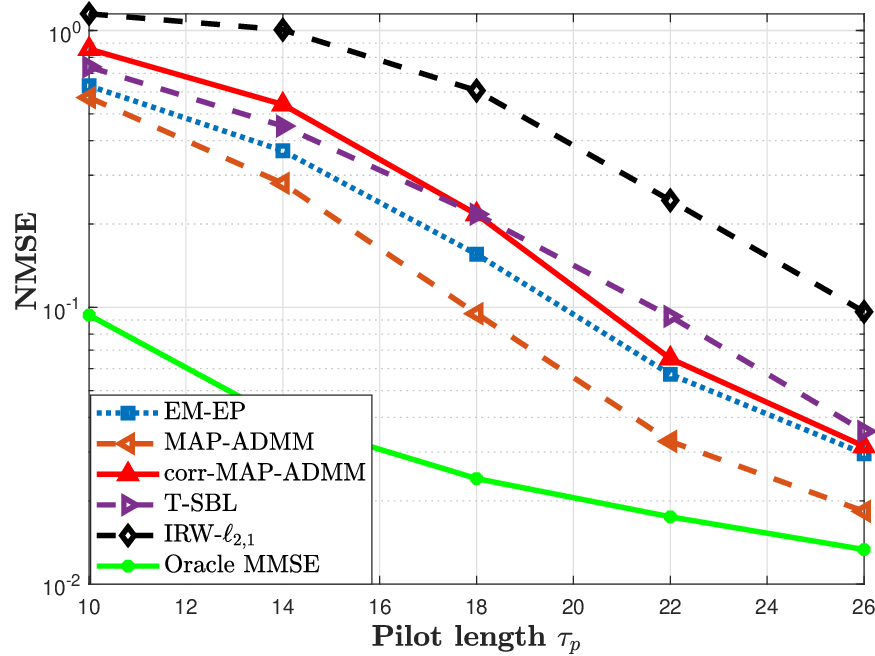}   \caption{}
     \label{mse_clus}
\end{subfigure}\vspace{-3mm}
\caption{JUICE Performance in independent activity pattern  in terms of: (a) SRR, (b) NMSE. }
\label{fig:results_rand}
\vspace{-5mm}
\end{figure}

We investigate now the robustness of the proposed algorithms in  scenarios of  sparsity model mismatch  where discrepancies exist between the actual scenario and the imposed algorithm model. A sparsity model mismatch  occurs if the UEs activity is independent  rather than exhibiting  a correlated pattern. 
Fig.\ \ref{fig:results_rand}(a)  illustrates the  SRR  against the pilot length for the different algorithms in \emph{ an independent activity pattern}. The results show that  EM-EP and corr-MAP-ADMM clearly outperform IRW-$\ell_{2,1}$ and  provide a slight improvement over T-SBL  and match MAP-ADMM performance for $\tau_{\mathrm{p}}>20$.
Furthermore, Fig.\  \ref{fig:results_rand}(b) shows that while the proposed algorithms outperform IR-$\ell_{2,1}$,   EM-EP   is slightly inferior  to MAP-ADMM, and it outperforms T-SBL in terms of NMSE, whereas, T-SBL provides the same performance as corr-MAP-ADMM.

The results shown in Fig.  \ref{fig:results_rand} 
demonstrate that  the proposed algorithms  yield comparable results to state-of-the-art algorithms, even in an activity pattern model mismatch and  no prior knowledge on the CDI.  More precisely, even though the proposed \emph{strong} hierarchical spike-and-slab \eqref{spike-slab} and the \emph{weak}  log-sum-based cluster-sparsity \eqref{eq:T2} 
 priors  were designed to favour clustered sparse solution, they are still flexible and \emph{robust} to  changes in the underlying sparsity structure.



\subsection{Convergence Behaviour}
\begin{figure}[ht]
    \centering
    \includegraphics[scale=.28]{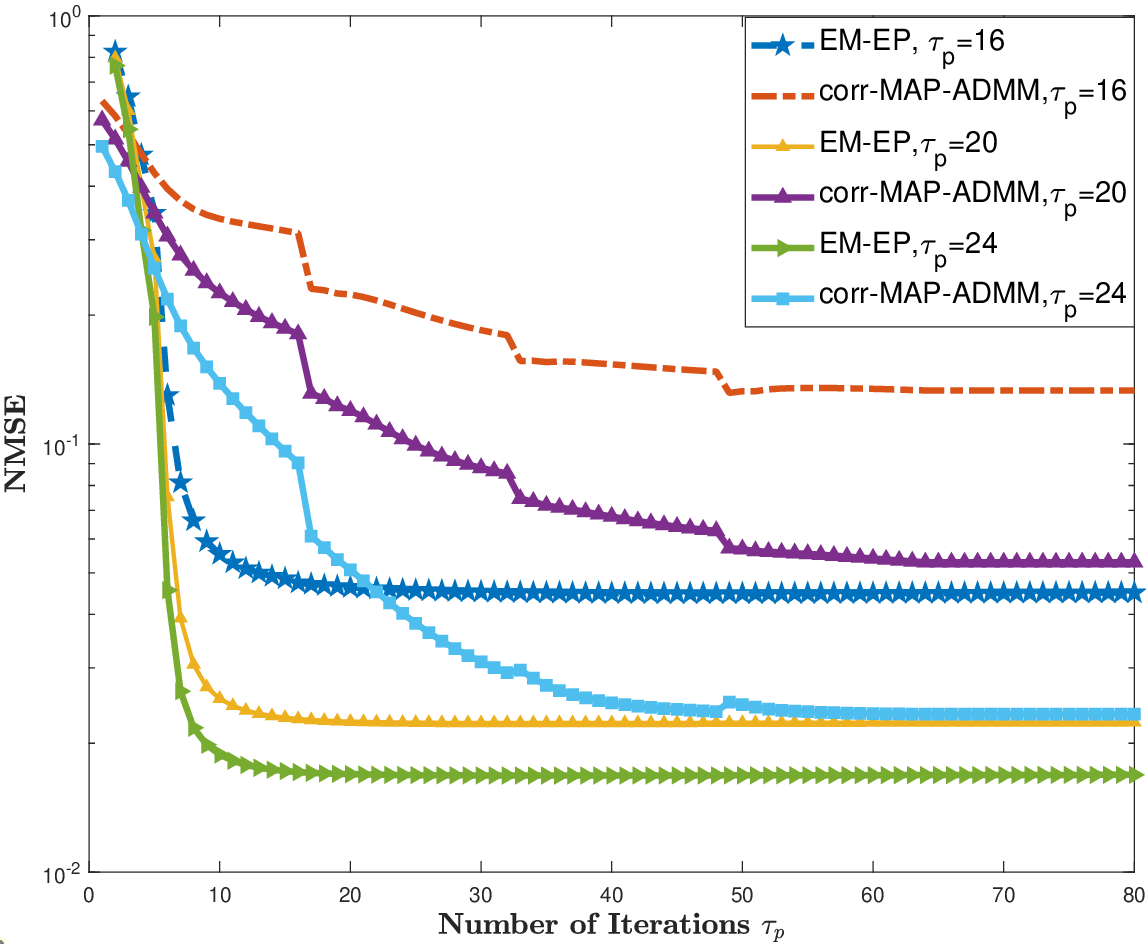}
    \caption{NMSE versus number of iterations for the proposed algorithms }
    \label{fig:convergence}
\end{figure}

The computational complexity per iteration of EM-EP is mainly dominated by the $\Sigmab$ update in \eqref{eq:Q_moments} and requires $\mathcal{O}(M^3N^3)$ complex multiplications. In contrast, corr-MAP-ADMM's primary computational complexity arises from the $\vec{V}$ and the $\vec{R}$ updates resulting in computational complexity in the order of  $\mathcal{O}(NM^2+N_{\mathrm{c}}M^3)$. For a reference,  MAP-ADMM and T-SBL require $\mathcal{O}(MN^2+NM^2)$ and $\mathcal{O}(N^2M^3\tau_{\mathrm{p}})$ complex multiplications, respectively. Nonetheless, despite being more computationally demanding compared to corr-MAP-ADMM, EM-EP exhibits a faster convergence rate, requiring only about 10 iterations to converge in contrast to corr-MAP-ADMM which takes up to 60 iterations as shown in Fig. \ref{fig:convergence}. 
\vspace{-1cm}\section{Conclusions  and Future Work}
\label{conclusion}
We addressed the JUICE under spatially correlated channels with a structured UEs activity constraints. In order to tackle this problem, we have developed two algorithms from  a Bayesian perspective by utilizing appropriate structured sparsity-promoting priors to capture both cluster-level and intra-cluster sparsity.
We numerically showed that the proposed algorithms provide a huge gain compared to state-of the art algorithms as well as exhibiting a strong robustness against activity pattern model mismatch. 

A potential future work would be dedicated toward  designing a distributed-type EM-EP algorithm in order to reduce the computational complexity of the EM-EP. This can be achieved for instance in cell-free MTC networks, where the computation load is distributed over several access points. A second line of research is to quantify the quality of the estimated covariance matrices to be  utilized, for example, in MMSE channel estimation.
\vspace{-.3cm}

\appendix
\subsection{ Product and Quotient Rules}\label{app:2gauss}
This section derives the product and quotient rules of two Gaussian random vector $\vecgreek{\theta}_1$ and $\vecgreek{\theta}_2$.
\subsubsection{Product}
\begin{equation}\label{eq:prod}
   \mathcal{N}(\vecgreek{\theta}_1;\vecgreek{\mu}_1,\Sigmab_1)\mathcal{N}(\vecgreek{\theta}_2;\vecgreek{\mu}_2,\Sigmab_2)=  K_{\mathrm{p}}\mathcal{N}(\vecgreek{\theta}_{\mathrm{p}};\vecgreek{\mu}_{\mathrm{p}},\Sigmab_{\mathrm{p}}),
\end{equation}
where $\Sigmab_{\mathrm{p}}=\big(\Sigmab_1^{-1}+\Sigmab_2^{-1}\big)^{-1}$, $\vec{\mu}_{\mathrm{p}}=\Sigmab^{-1}\big(\Sigmab_1^{-1}\vecgreek{\mu}_1+\Sigmab_2^{-1}\vecgreek{\mu}_2 \big)$, and $K_{\mathrm{p}}={\mathcal{N}(\vecgreek{\mu}_1;\vecgreek{\mu}_2, \Sigmab_2+\Sigmab_1)}$.
 \subsubsection{Fraction}
\begin{equation}\label{eq:frac}
   \frac{\mathcal{N}(\vecgreek{\theta}_1;\vecgreek{\mu}_1,\Sigmab_1)}{\mathcal{N}(\vecgreek{\theta}_2;\vecgreek{\mu}_2,\Sigmab_2)}=  K_{\mathrm{q}}\mathcal{N}(\vecgreek{\theta}_{\mathrm{q}};\vecgreek{\mu}_{\mathrm{q}},\Sigmab_{\mathrm{q}}), 
\end{equation}
where $\Sigmab_{\mathrm{q}}=\big(\Sigmab_1^{-1}-\Sigmab_2^{-1}\big)^{-1}$, $\vec{\mu}=\Sigmab^{-1}\big(\Sigmab_1^{-1}\vecgreek{\mu}_1-\Sigmab_2^{-1}\vecgreek{\mu}_2 \big)$, and  $K_{\mathrm{q}}=\frac{|\Sigmab_2\Sigmab_1^{-1}|}{\mathcal{N}(\vecgreek{\mu}_1;\vecgreek{\mu}_2, \Sigmab_2-\Sigmab_1)}$.

Thus, by using \eqref{eq:prod} and \eqref{eq:frac}, and few simple manipulations,  we obtain straightforward the terms in \eqref{eq:Q},\eqref{Q_cavity}, \eqref{eq:q_new}.\vspace{-.6cm}
\subsection{Derivation of  $q_2(\cdot)$ Update} \label{app::q_2}
We start by computing the normalizing constant for $f_2(\vec{X}_{\mathcal{C}_l},c_l)Q\ex{2,l}(\vec{X}_{\mathcal{C}_l},c_l)$ 
\begin{equation}\footnotesize\label{eq:Gl,0}
\begin{array}{ll}
       G_{l,0}&=\displaystyle\sum_{c_l}\int  f_2(\vec{X}_{\mathcal{C}_l},c_l)Q\ex{2,l}(\vec{X}_{\mathcal{C}_l},c_l)d\vec{X}_{\mathcal{C}_l}dc_l \\
    
    &\overset{(a)}{=}\displaystyle\sum_{c_l}\int  \big[(1-c_l)\prod_{i\in \mathcal{C}_l}\delta\big(\vec{x}_i\big)+c_l\prod_{i\in \mathcal{C}_l} \mathcal{CN}(\vec{x}_i;\vec{0},\vec{R}_i ) \big]\mathcal{CN}(\vec{x}_i;\hat{\vec{m}}_i\ex{2,l},\hat{\Sigmab}_i\ex{2,l})\mathcal{B}(c_l|\epsilon)d\vec{X}_{\mathcal{C}_l}\\
    
  &=\displaystyle (1-\epsilon) \int\prod_{i\in \mathcal{C}_l}\delta\big(\vec{x}_i\big)\mathcal{CN}(\vec{x}_i;\hat{\vec{m}}_i\ex{2,l},\Sigmab_i\ex{2,l})d\vec{X}_{\mathcal{C}_l}\\&+\displaystyle\epsilon \int  \mathcal{CN}(\vec{x}_i;\vec{0},\vec{R}_i) \mathcal{CN}(\vec{x}_i;\hat{\vec{m}}_i\ex{2,l},\hat{\Sigmab}_i\ex{2,l})d\vec{x}_i\prod_{j\notin \mathcal{C}_l} \int \mathcal{CN}(\vec{x}_j;\vec{0},\vec{R}_j) \mathcal{CN}(\vec{x}_j;\hat{\vec{m}}_j\ex{2,l},\hat{\Sigmab}_j\ex{2,l})d\vec{x}_j\\
  
      &\displaystyle\overset{(b)}{=} (1-\epsilon) \prod_{i \in \mathcal{C}_l}\mathcal{CN}(\vec{0};\hat{\vec{m}}_i\ex{2,l},\hat{\Sigmab}_i\ex{2,l})   +\epsilon\displaystyle\int \prod_{i \in \mathcal{C}_l} \underbrace{  \mathcal{CN}(\vec{x}_i;\vec{R}_i(\vec{R}_i+\hat{\Sigmab}_i\ex{2,l})^{-1}\hat{\vec{m}}_i\ex{2,l} ,\big(\vec{R}_i^{-1}+\hat{\Sigmab}_i\ex{2,l^{-1}})^{-1}\big)d\vec{x}_i}_{\overset{(c)}{=}1} \\ &\displaystyle \prod_{i \in \mathcal{C}_l}\mathcal{CN}(\vec{0};\hat{\vec{m}}_i\ex{2,l},\hat{\Sigmab}_i\ex{2,l}+\vec{R}_i)\overset{(d)}{=}b_l+ a_l,\\
      
\end{array}
\end{equation}
where (a) is obtained by plugging the densities, (b) by straightforward application of  \eqref{eq:prod}, (c) by integrating all the involved Gaussian  distributions to unity, and (d) by setting  $b_l=(1-\epsilon)\displaystyle  \prod_{i \in \mathcal{C}_l}\mathcal{CN}(\vec{0};\hat{\vec{m}}_i\ex{2,l},\hat{\Sigmab}_i\ex{2,l}) $ and ${a_l=\epsilon\displaystyle\prod_{i \in \mathcal{C}_l}\mathcal{CN}(\vec{0};\hat{\vec{m}}_i\ex{2,l},\hat{\Sigmab}_i\ex{2,l}+\vec{R}_i)}$. 

Subsequently, by following the same steps, the $n$th moment with respect to $\vec{x}_i$ is computed as \begin{equation}\footnotesize\label{eq:nth_momement}
\begin{array}{ll}
       \mathbb{E}_{f_2Q\ex{2,l}}[\vec{x}_i^{n}]&=\frac{1}{G_{l,0}}\displaystyle\sum_{c_l}\int \vec{x}_i^{n} f_2(\vec{X}_{\mathcal{C}_l},c_l)Q\ex{2,l}(\vec{X}_{\mathcal{C}_l},c_l)d\vec{X}_{\mathcal{C}_l}dc_l \\

    &{=}\frac{a_l}{G_{l,0}}\displaystyle\int \vec{x}_i^{n} \mathcal{CN}(\vec{x}_i;\vec{R}_i(\vec{R}_i+\hat{\Sigmab}_i\ex{2,l})^{-1}\hat{\vec{m}}_i\ex{2,l} ,\big(\vec{R}_i^{-1}+\hat{\Sigmab}_i\ex{2,l^{-1}})^{-1}\big)d\vec{x}_i,
\end{array}
\end{equation}
Note that the  integral term represents the $n$th moment of a multivariate Gaussian random vector, thus the mean $\mathbb{E}_{f_2Q\ex{2,l}}[\vec{x}_i]$ and the variance $\textrm{Var}_{f_2Q\ex{2,l}}[\vec{x}_i]$ are deduced directly from \eqref{eq:nth_momement} and they are given in  \eqref{q_new_moments}. 
Finally, the posterior mean with respect to $c_l$  is computed as
\begin{equation}\footnotesize
\begin{array}{ll}
  \mathbb{E}_{f_2Q\ex{2,l}}[c_l]&= \frac{1}{G_{l,0}}\displaystyle\sum_{\vec{c}}\int c_l f_2(\vec{X}_{\mathcal{C}_l},c_l)Q\ex{2,l}(\vec{X}_{\mathcal{C}_l},\vec{c})d\vec{X}d\vec{c}  \\

&\overset{}{=}\displaystyle\sum_{c_l} c_l (1-c_l)\mathcal{B}(c_l;\epsilon)dc_l\int\prod_{i\in \mathcal{C}_l}\delta\big(\vec{x}_i\big)\mathcal{CN}(\vec{x}_i;\hat{\vec{m}}_i\ex{2,l},\hat{\Sigmab}_i\ex{2,l})\\&+\sum_{c_l}c_l^2\mathcal{B}(c_l;\epsilon)dc_l\int\prod_{i\in \mathcal{C}_l} \mathcal{CN}(\vec{x}_i;\vec{0},\vec{R}_i ) \mathcal{CN}(\vec{x}_i;\hat{\vec{m}}_i\ex{2,l},\hat{\Sigmab}_i\ex{2,l}) d\vec{X}_{\mathcal{C}_l}\\

   &\overset{}{=}\frac{1}{G_{l,0}}\mathbb{E}[c_l-c_l^2]\displaystyle  \prod_{i \in \mathcal{C}_l}\mathcal{CN}(\vec{0};\hat{\vec{m}}_i\ex{2,l},\hat{\Sigmab}_i\ex{2,l}) +\frac{1}{G_{l,0}}\mathbb{E}[c_l^2]\prod_{i \in \mathcal{C}_l}\mathcal{CN}(\vec{0};\hat{\vec{m}}_i\ex{2,l},\hat{\Sigmab}_i\ex{2,l}+\vec{R}_i)=\frac{a_l}{G_{l,0}}.

    
\end{array}
\end{equation}

\bibliographystyle{IEEEtran}
\mybibliography

\end{document}